%% file: main.tex
\documentclass[a4paper,journal]{IEEEtran}
\IEEEoverridecommandlockouts

\input{settings/packages}

\input{settings/macros}
\input{settings/commands}
\input{settings/acronyms}

\usepackage{fancyhdr}
\fancypagestyle{IEEEtitlepagestyle}{
    \fancyhf{} 
     
    \fancyhead[C]{
        \small This work has been submitted for possible publication.
        }
}

\begin{document}

\title{DPDS: A DPDK-Based Packet Delayer and Spacer}

\author{
  \IEEEauthorblockN{Etienne Zink$^{\orcidicon{0009-0001-0879-535X}}$, Fabian Ihle$^{\orcidicon{0009-0005-3917-2402
        }}$,  and Michael Menth$^{\orcidicon{0000-0002-3216-1015}}$}\\
  \IEEEauthorblockA{University of T\"ubingen, Chair of Communication Networks
    \\Email: \{etienne.zink, fabian.ihle, menth\}@uni-tuebingen.de}
  \thanks{The authors acknowledge funding in part by the Deutsche Forschungsgemeinschaft (DFG) under grant 503231190, and in part by the Open Access Publishing Fund of the University of T\"ubingen. Furthermore, the authors acknowledge the use of Claude Opus 4.8 to assist in developing scripts for experiment automation and data visualization presented in this paper.}%
}


\maketitle

\begin{abstract}
  In this paper we tackle the problem of adding varying delay to packets for link emulation.
  Naive approaches either add more delay than desired or cause packet reordering, both of which are undesirable.
  We develop adaptive delay correlation, which adds positively correlated delays to packets efficiently.
  It takes a mean delay and standard deviation (jitter) as input, as well as a half-life period to control the delay dynamics.
  We investigate the accuracy and dynamics of the resulting packet delays with and without bandwidth limitation.
  As a result we give a recommendation for the configuration of the half-life period.
  We implement adaptive delay correlation in a DPDK-based packet delayer and spacer (DPDS), investigate its performance on hardware, and compare it with the widely used link emulator NetEm and the recently developed DPDK-based emulator MoonEm.
  DPDS outperforms both of them with a zero-loss throughput of 95 Gbit/s for constant delay and, with spacing enabled, 85 Gbit/s for varying delay with 3 ms jitter.
  Further, DPDS supports packet reordering with zero-loss throughputs of 73 Gbit/s and 58 Gbit/s for constant and varying delay, respectively, as well as policing and two packet loss models.
\end{abstract}


\input{content/introduction}
\input{content/relatedWork}
\input{content/theory}
\input{content/simulation}
\input{content/dpds}
\input{content/evaluation}
\input{content/conclusions}

{
  \acsetup{
    list/template = tabular,
    templates/colspec = {>{\bfseries}l@{\hspace{0.5em}}p{0.78\columnwidth}},
    list/preamble = \vspace{\baselineskip}
  }
  \printacronyms[name=List of Abbreviations]
  \vspace{\baselineskip}
}

\bibliographystyle{IEEEtran}
\bibliography{bibliography/conferences, bibliography/literature}

\input{content/authors}

\end{document}

%% file: settings/packages.tex
\usepackage{cite}
\usepackage{graphicx}
\usepackage{amsmath,amssymb,amsfonts}
\usepackage{xcolor}
\usepackage{xfrac}
\usepackage{algorithmic}
\usepackage{textcomp}
\usepackage{acro}
\usepackage{xurl}
\usepackage{siunitx}
\usepackage{booktabs}
\usepackage{tabularx}
\usepackage{multirow}
\usepackage{xspace}
\usepackage{csquotes}
\usepackage{hyperref}
\usepackage[caption=false]{subfig}
\usepackage{scalerel}
\usepackage{tikz}
\usepackage{makecell}
\usepackage[english]{babel}
\usepackage{fontawesome5}
\usepackage{mathtools}

%% file: settings/macros.tex
\newcommand\citeN\cite

\newboolean{makevspace}
\newcommand{\cvspace}[1]{%
   \ifthenelse
   {\boolean{makevspace}}
   {\vspace{#1}}
   {}%
}

%% file: settings/commands.tex
\let\OldTexttrademark\texttrademark
\renewcommand{\texttrademark}{\OldTexttrademark\xspace}

\DeclareSIUnit{\bps}{bit/s}
\DeclareSIUnit{\kbps}{kbit/s}
\DeclareSIUnit{\mbps}{Mbit/s}
\DeclareSIUnit{\gbps}{Gbit/s}
\DeclareSIUnit{\tbps}{Tbit/s}
\DeclareSIUnit{\B}{B}
\DeclareSIUnit{\kB}{kB}
\DeclareSIUnit{\mB}{MB}
\DeclareSIUnit{\gB}{GB}
\DeclareSIUnit{\tB}{TB}
\DeclareSIUnit{\b}{b}
\DeclareSIUnit{\kb}{kb}
\DeclareSIUnit{\mb}{Mb}
\DeclareSIUnit{\gb}{Gb}
\DeclareSIUnit{\tb}{Tb}
\DeclareSIUnit{\pps}{pps}
\DeclareSIUnit{\kpps}{kpps}
\DeclareSIUnit{\us}{\micro\second}
\DeclareSIUnit{\ms}{\milli\second}
\DeclareSIUnit{\min}{\minute}

\newcolumntype{Y}{>{\centering\arraybackslash}X}
\def\BibTeX{{\rm B\kern-.05em{\sc i\kern-.025em b}\kern-.08em
        T\kern-.1667em\lower.7ex\hbox{E}\kern-.125emX}}

\usetikzlibrary{svg.path}
\definecolor{orcidlogocol}{HTML}{A6CE39}
\tikzset{
    orcidlogo/.pic={
            \fill[orcidlogocol] svg{M256,128c0,70.7-57.3,128-128,128C57.3,256,0,198.7,0,128C0,57.3,57.3,0,128,0C198.7,0,256,57.3,256,128z};
            \fill[white] svg{M86.3,186.2H70.9V79.1h15.4v48.4V186.2z}
            svg{M108.9,79.1h41.6c39.6,0,57,28.3,57,53.6c0,27.5-21.5,53.6-56.8,53.6h-41.8V79.1z M124.3,172.4h24.5c34.9,0,42.9-26.5,42.9-39.7c0-21.5-13.7-39.7-43.7-39.7h-23.7V172.4z}
            svg{M88.7,56.8c0,5.5-4.5,10.1-10.1,10.1c-5.6,0-10.1-4.6-10.1-10.1c0-5.6,4.5-10.1,10.1-10.1C84.2,46.7,88.7,51.3,88.7,56.8z};
        }
}
\newcommand\orcidicon[1]{\href{https://orcid.org/#1}{%
        \mbox{\begin{tikzpicture}[yscale=-1,transform shape, scale=0.03] 
                \pic{orcidlogo};
            \end{tikzpicture}}}}

\hypersetup{
    colorlinks = true,
    allcolors=blue,
    urlcolor=black,
    linkcolor=black
}

\newcommand*\circled[1]{\tikz[baseline=(char.base)]{
        \node[shape=circle,fill=black,text=white,inner sep=0.5pt] (char) {#1};}}

\addto\extrasenglish{
    
}

\addto\extrasenglish{
    
}

\addto\extrasenglish{
    
}

\addto\extrasenglish{
    
}

\addto\extrasenglish{
    
}

%% file: settings/acronyms.tex
\acsetup{single}

\DeclareAcronym{RTT}{
    short = RTT,
    long = round trip time,
}

\DeclareAcronym{NIC}{
    short = NIC,
    long = network interface card,
}

\DeclareAcronym{IAT}{
    short = IAT,
    long = inter-arrival time,
}

\DeclareAcronym{TBF}{
    short = TBF,
    long = token bucket filter,
}

\DeclareAcronym{DPDS}{
    short = DPDS,
    long = {DPDK-based packet delayer and spacer},
}

\DeclareAcronym{FIFO}{
    short = FIFO,
    long = {first in, first out},
}

\DeclareAcronym{STD}{
    short = STD,
    long = standard deviation,
}

\DeclareAcronym{QDisc}{
    short = QDisc,
    long = queueing discipline,
}

\DeclareAcronym{TSN}{
    short = TSN,
    long = time-sensitive networking,
    single = TSN,
}

\DeclareAcronym{VNE}{
    short = VNE,
    long = virtual network emulator,
}

\DeclareAcronym{NLE}{
    short = NLE,
    long = network link emulator,
}

\DeclareAcronym{DMA}{
    short = DMA,
    long = direct memory access,
}

\DeclareAcronym{EMA}{
    short = EMA,
    long = exponential moving average,
}

\DeclareAcronym{CBR}{
    short = CBR,
    long = constant bit-rate,
}

\DeclareAcronym{DES}{
    short = DES,
    long = discrete event simulation,
}

\DeclareAcronym{ZLT}{
    short = ZLT,
    long = zero-loss throughput,
    long-plural-form = zero-loss throughputs,
}

\DeclareAcronym{OOO}{
    short = OOO,
    long = out-of-order,
}

%% file: content/introduction.tex
\section{Introduction}
\label{sec:introduction}

\Acp{NLE} are often needed for performance studies of network protocols.
They typically delay packets according to a specified mean delay and standard deviation (jitter), limit bandwidth using a spacer, police packets exceeding a traffic contract, and drop packets according to some loss model.
As network speeds keep increasing, \acp{NLE} that support high traffic rates are also needed.
While fast \acp{NLE} add only constant packet delays, some studies point out that packet delay over networks typically varies~\cite{Pa99,HoAh16}.
However, there is currently no high-performance \ac{NLE} that supports varying delay appropriately.

Adding varying delay is harder than it appears.
If the added delay for a packet is substantially longer than the delay for its successor, the successor packet will face additional delay, leading to achieved mean delays that are longer than desired.
Packet reordering according to transmission times after adding delay solves the problem, but packet reordering is detrimental to many transport or application protocols, such as TCP~\cite{BlAl02,ZhKa02,LaGe02}, QUIC~\cite{KaJe17}, and RDMA~\cite{MiSh18,SoKh23}.
An alternative to reordering is smoothing consecutive delays with an \ac{EMA} so that the resulting packet delays are correlated.
However, the configuration of the \ac{EMA}'s internal weight is left to the user and the weight does not adapt to the current traffic rate.

In this paper we develop an efficient algorithm that adds varying delay to packets using the \ac{EMA} approach.
The algorithm is configured with a desired mean delay and jitter, and a half-life period to control the delay dynamics.
Moreover, the algorithm continuously measures the current packet rate and adapts the \ac{EMA}'s internal weight according to the configuration parameters and the measured packet rate.
We further investigate the performance of uncorrelated delays, adaptive delay correlation, and packet reordering using simulation with and without bandwidth limitations.
The results give insights into the accuracy of the different approaches and lead to a recommendation for the configuration of the half-life period.
Building on these insights, we implement adaptive delay correlation in a DPDK-based \ac{NLE}, called \ac{DPDS}.
We evaluate its \ac{ZLT} and maximum supported rate and compare them with those of the widely used \ac{NLE} NetEm and the recently developed DPDK-based emulator MoonEm.
We show that the desired delay is well achieved under \ac{ZLT}, and that \ac{DPDS} also works as intended under varying traffic rates, which is challenging as its internal operation depends on measured traffic rates.

The remainder of this paper is structured as follows:
\autoref{sec:relatedWork} reviews related work.
\autoref{sec:adaptiveCorrelatedPacketDelays} introduces adaptive delay correlation.
\autoref{sec:simulation} evaluates adaptive delay correlation in simulation.
\autoref{sec:theDPDSEmulator} describes the emulator \ac{DPDS}.
\autoref{sec:prototypeEvaluation} evaluates the \ac{DPDS} prototype by experimentation.
\autoref{sec:conclusions} concludes the paper.

%% file: content/relatedWork.tex
\section{Related Work}
\label{sec:relatedWork}

Nussbaum and Richard~\cite{NuRi09} distinguish two concepts of network emulators: \acfp{VNE} and \acfp{NLE}.
\acp{VNE} emulate a whole network.
Examples of \acp{VNE} are Mininet~\cite{HaHe12}, its extension Containernet 2.0~\cite{PeKa18}, and Kollaps~\cite{GoNe20}.
In contrast, \acp{NLE} apply link characteristics to packets as they are transmitted or received on an interface.
The emulated conditions include packet delay, limited bandwidth, and packet loss~\cite{GoKf23}.
Examples of \acp{NLE} are Dummynet~\cite{Ri97}, NetEm~\cite{He05}, DEMU~\cite{AkHi17}, and MoonEm~\cite{LaGa25}.
In this paper, we focus on \acp{NLE}.
We therefore use the terms \ac{NLE} and emulator interchangeably.

\renewcommand{\cellalign}{cc}
\begin{table}[t]
  \caption{Delay capabilities of existing emulators and \ac{DPDS}. \emph{Configurable}, \emph{Fixed}, and \emph{Adaptive} indicate whether the correlation weight is user-selectable, fixed by the implementation, or recomputed from the current rate.}
  \label{tab:emulatorDelayCapabilities}
  \centering
  \begin{tabularx}{\columnwidth}{l|YYY}
    \toprule
                                     & \emph{Constant delay} & \multicolumn{2}{c}{\emph{Varying delay}}                      \\
    \cmidrule(lr){3-4}
                                     &                       & \emph{Reordering}                        & \emph{Correlation} \\\midrule
    \multicolumn{4}{l}{\textit{In-kernel}}                                                                                   \\\midrule
    Dummynet~\cite{Ri97}             & \checkmark            & $\times$                                 & $\times$           \\
    NetEm~\cite{He05}                & \checkmark            & \checkmark                               & Configurable       \\
    eBPF + QDisc~\cite{BePf22}       & \checkmark            & $\times$                                 & $\times$           \\
    6GDetCom\_Emulator~\cite{HaDu25} & \checkmark            & \checkmark                               & $\times$           \\
    Rattan~\cite{WaSh25}             & \checkmark            & $\times$                                 & $\times$           \\
    TheaterQ~\cite{OtHi25}           & \checkmark            & \checkmark                               & $\times$           \\\midrule
    \multicolumn{4}{l}{\textit{Kernel bypass}}                                                                               \\\midrule
    DEMU~\cite{AkHi17}               & \checkmark            & $\times$                                 & $\times$           \\
    MoonGen LTE~\cite{StWa20}        & \checkmark            & $\times$                                 & $\times$           \\
    MoonEm~\cite{LaGa25}             & \checkmark            & $\times$                                 & $\times$           \\
    SmartNet~\cite{VoRo25}           & \checkmark            & $\times$                                 & Fixed              \\
    \acs{DPDS}                       & \checkmark            & \checkmark                               & Adaptive           \\
    \bottomrule
  \end{tabularx}
\end{table}

We classify \acp{NLE} into two categories: emulators implemented in the Linux kernel and emulators implemented with kernel bypass frameworks.
\autoref{tab:emulatorDelayCapabilities} presents an overview of \acp{NLE} of both classes.
It further distinguishes whether the emulators support constant delays, i.e., a jitter/\ac{STD} of zero, or varying delays, i.e., a jitter greater than zero.
To accurately emulate varying delays, emulators implement either packet reordering (sorting packets according to their transmission times) or delay correlation.
Although reordering can accurately emulate varying delays, it impairs in-order delivery.

\subsection{In-Kernel Emulators}
\label{subsec:inKernelEmulators}

Dummynet and NetEm are the most commonly used emulators in academic research~\cite{StWa20} and have been thoroughly evaluated in several studies~\cite{NuRi09,LuBu14,StWa20,GoKf23}.
Rizzo~\cite{Ri97} introduced Dummynet in 1997, and Carbone and Rizzo~\cite{CaRi10} enhanced it further in 2010.
Dummynet emulates bandwidth limitation and constant delays on a single workstation.
NetEm was introduced by Hemminger~\cite{He05} in 2005 to be more extensible than Dummynet.
It is implemented as a \ac{QDisc} in Linux's traffic control, and can be combined with other \acp{QDisc}, like the \ac{TBF}~\cite{KuHu01} to implement bandwidth limitation.
NetEm supports delay emulation with reordering and correlated delays.
It implements delay correlation through an \ac{EMA} with a configurable correlation weight.
The correlation weight determines how similar consecutive delays are.
However, according to Hemminger~\cite{He05}, NetEm's correlated delays are inaccurate.

Later, additional in-kernel emulators were developed to overcome the limitations of Dummynet and NetEm.
Becker et al.~\cite{BePf22} implemented a constant delay emulator based on eBPF and a Linux \ac{QDisc} to increase scalability in large-scale virtual edge testbeds.
Haug et al.~\cite{HaDu25} introduced the 6GDetCom\_Emulator, a custom Linux \ac{QDisc}, which emulates varying delays of 5G TSN bridges with packet reordering at rates of up to \qty{1}{\gbps}~\cite{Slides6GDetCom_Emulator}.
Wang et al.~\cite{WaSh25} developed Rattan, a Rust-based constant delay emulator.
However, Rattan has not yet been thoroughly evaluated.
Recently, Ottens et al.~\cite{OtHi25} introduced TheaterQ, a custom Linux \ac{QDisc}, to emulate varying delay with reordering~\cite{GitHubTheaterQ}.
Of these four emulators, only two, i.e., 6GDetCom\_Emulator and TheaterQ, support varying delays with reordering.
Neither supports correlated delays.

\subsection{Kernel Bypass-Based Emulators}
\label{subsec:kernelBypassEmulator}

Emulators based on kernel bypass frameworks were developed to surpass the accuracy and throughput of existing in-kernel emulators~\cite{AkHi17,LaGa25}.
Kernel bypass frameworks, like DPDK~\cite{DPDKWebsite} and Snabb~\cite{PaNi15}, leverage \ac{DMA} to completely bypass the Linux kernel networking stack, and pass packets from the \ac{NIC} directly to the user space application~\cite{EmPu19}.
This reduces the overhead of the generalized Linux networking stack.

Three DPDK-based \acp{NLE} have been proposed: DEMU~\cite{AkHi17}, MoonGen LTE~\cite{StWa20}, and MoonEm~\cite{LaGa25}.
DEMU, developed by Aketa et al.~\cite{AkHi17}, was the first to accurately emulate short (sub-millisecond) constant delays.
It was later enhanced by Sasaki et al.~\cite{SaHi19} to emulate packet loss and by Puakalong et al.~\cite{PuTa20} to emulate bandwidth limitations.
Stratmann et al.~\cite{StWa20} and Lachnit et al.~\cite{LaGa25} developed two emulators based on the MoonGen~\cite{EmGa15} traffic generator.
Stratmann et al.~\cite{StWa20} developed an emulation of various constant delays and bandwidths for the uplink and downlink of an LTE testbed.
Lachnit et al.~\cite{LaGa25} introduced MoonEm, an extension to MoonGen which uses NIC hardware timestamping to accurately emulate constant delays.
None of these three emulators support varying delays.

In contrast, Vogt et al.~\cite{VoRo25} demonstrated SmartNet, a DPDK-based \ac{VNE} for SmartNIC-aware network emulation.
SmartNet primarily focuses on emulating a virtual network, but can also emulate various link characteristics, such as varying delays, inside this virtual network.
It supports correlated delays through an \ac{EMA} with a fixed correlation weight~\cite{GitHubSmartNet}.

%% file: content/theory.tex
\section{Adaptive Correlated Packet Delays}
\label{sec:adaptiveCorrelatedPacketDelays}

In the following we develop a method to add correlated delay values to consecutive packets.
It greatly diminishes the drawbacks of independent delays, which are a larger mean delay and lower jitter than desired.
The method is based on the \ac{EMA} that generates delay values $C_i$ and averages them to delay values $D_i$, which are used for delaying the packets.

We first derive appropriate parameters for $C$ to meet a desired mean and jitter for $D$.
This approach depends on a correlation weight $\alpha$, which controls its dynamics.
Then, we make the technique adaptive to different packet rates $R$ by setting $\alpha$ such that a desired half-life period $t_h$ is met for generated delay values $C_i$.
Finally, we make it adaptive to changing packet rates $R$ by measuring the rate on the fly.

\subsection{Exponentially Weighted Moving Average for Correlated Packet Delays}

We successively generate delays $C_i$ and average them with a correlation weight parameter $\alpha$ by
\begin{align*}
    D_i=\alpha\cdot D_{i-1}+(1-\alpha)\cdot C_i
\end{align*}
for $i\geq 1$ and $D_0=C_0$.
This is an \ac{EMA}, background information is available in \cite{MeHa17}.
The generating delays $C_i$ are independent and identically distributed random variables, $C$ for short.
Therefore, the averaged delays $D_i$ have the same mean as the generated delays $C_i$.
The variance of $D_i$ can be computed as
\begin{align*}
    VAR[D_i] & = VAR[\alpha^i \cdot C_0 + \sum_{0\le j<i}(1-\alpha)\cdot \alpha^j \cdot C_{i-j}] \\
             & = \alpha^{2\cdot i} \cdot VAR[C_0]                                                \\
             & + \sum_{0\le j<i}((1-\alpha)\cdot\alpha^j)^2 \cdot VAR[C_{i-j}].
\end{align*}

In the limit for $i\to\infty$ and for $0 < \alpha < 1$, we obtain
\begin{align*}
    \sigma^2 = VAR[D] = \lim_{i\to\infty} VAR[D_i] & = \frac{1-\alpha}{1+\alpha}\cdot VAR[C].
\end{align*}
Thus, to obtain a desired mean $\mu$ and jitter $\sigma$ for correlated delays $D_i$, the generating delay's mean should be set to $\mu$ and its variance to
\begin{align*}
    VAR[C]=\frac{1+\alpha}{1-\alpha}\cdot \sigma^2.
\end{align*}
While some \acp{NLE} offer the use of the \ac{EMA} for correlated delays \cite{He05,VoRo25}, none of them accounts for the derived relation between generating and averaged delays.

\subsection{Adapting the Dynamics to Different Rates}

The packet rate $R$ governs the dynamics of correlated delays, i.e., the speed at which the time series $D_i$ takes significantly different values over time.
We first assume a constant packet rate $R$.
A generating delay $C_i$ contributes to $D_{i+k}$ with weight $(1-\alpha)\cdot\alpha^k$, i.e., $C_i$ influences $D_{i+k}$ only $\alpha^k$ times as much as it influences $D_i$.
The half-life distance $k_h$ is the number of averaging steps after which this influence is at most halved, i.e.,
\begin{align*}
    k_h = \min\left\{\, k \in \mathbb{N},\ k > 0 \;:\; \alpha^k \le \tfrac{1}{2} \,\right\}.
\end{align*}
We want to configure a half-life period $t_h$, i.e., the time after which the impact of $C_i$ on the averaged delays is halved, which is $t_h=\frac{k_h}{R}$, so that $k_h=t_h\cdot R$.
This yields
\begin{align*}
    \alpha=\sqrt[t_h\cdot R]{\sfrac{1}{2}}.
\end{align*}
Thus, we propose to set the correlation weight $\alpha$ depending on a configured half-life period $t_h$ of the generating delays $C$ and the packet rate $R$.

\subsection{Adapting the Dynamics to Changing Rates}

As the packet rate may be unknown and change over time, it must be measured during emulation.
We measure it with $R=\frac{N}{T}$, where $N$ is the number of packets observed in an interval of duration $T$.
The intervals need to be successive and non-overlapping.
Whenever an interval closes, $R$ and the weight $\alpha$ are recomputed and applied to all subsequent packets until the next interval closes.
To yield accurate results, the interval should be short compared to the time scale on which the rate changes.

To keep the estimate well-defined, we count arriving packets and close the current interval with the first packet that arrives after a minimum duration $T_{min}$ has elapsed.
This packet simultaneously opens the next interval.
Since each interval is thus both opened and closed by a packet arrival, at least one packet is always counted, i.e., $N\geq 1$, and the measured rate $R=\frac{N}{T}$ with $T\geq T_{min}$ is strictly positive by construction.
Therefore, no special handling of zero rates is required.
We use a default of $T_{min}=\qty{10}{\ms}$ in our evaluations.

%% file: content/simulation.tex
\section{Simulation of Varying Delays}
\label{sec:simulation}

We evaluate the adaptive delay correlation using a \ac{DES} to avoid potential effects from a real-world implementation.
The \ac{DES} is available on GitHub~\cite{GitHubDPDSsimulation}.
In the simulation, \ac{CBR} traffic with 1518-byte packets is delayed by a delay $D$ derived with adaptive delay correlation and a normal distribution.
Unless stated otherwise, each experiment takes \qty{10}{\min} and is repeated ten times to compute confidence intervals with a confidence level of 95\%.
Most confidence intervals are so small that they collapse to a point and are barely visible.

We first describe the problem of naively implementing uncorrelated, non-reordered varying delays.
Afterward, we study the impact of the half-life period on delay dynamics and accuracy for a mean delay of \qty{10}{\ms}.
Then we generalize the findings to other delay values.
We compare correlated packet delays with packet reordering.
First, we consider unlimited bandwidth for the experiments, i.e., packets can be sent without queuing delay.
Second, we take the effect of limited bandwidth on accuracy into account.
Both consider correlated packet delays and packet reordering.
Finally, we investigate the influence of different underlying distributions on adaptive delay correlation's accuracy.

\subsection{Problem of Uncorrelated, Non-Reordered Varying Delays}
\label{subsec:problemVaryingPacketDelay}

The simplest approach to emulate varying packet delays is their generation without any correlation or reordering.
However, such uncorrelated delays are inaccurate: even though the delays are generated independently, they still influence each other.
If the added delay for a packet is substantially longer than the delay for its successor, the successor packet will face additional delay.
This additional queuing delay increases the achieved mean delay and reduces the achieved jitter.

\begin{figure}[t]
  \centering
  \subfloat[Mean delay accuracy.\label{fig:combined_delay_normal_uncorrelated}]{%
    \includegraphics[width=\columnwidth]{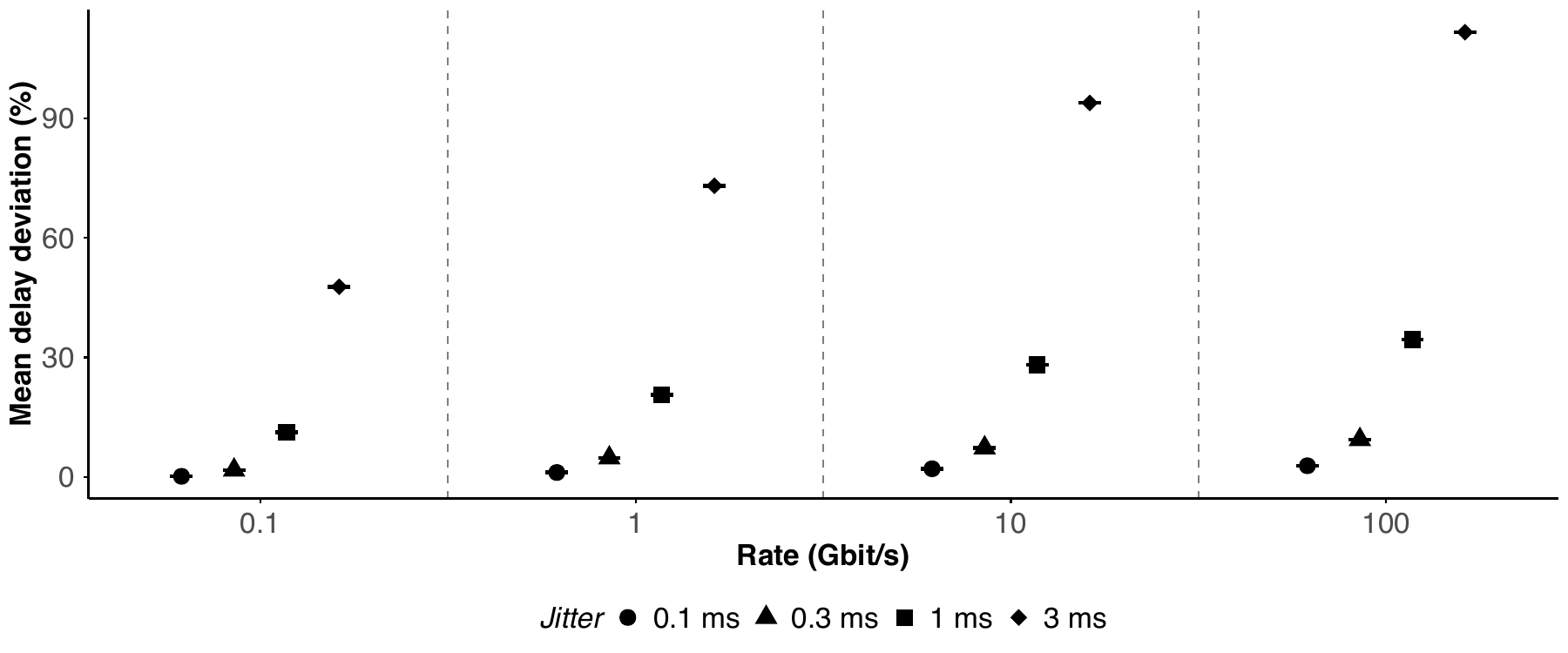}%
  }\\
  \subfloat[Jitter accuracy.\label{fig:combined_jitter_normal_uncorrelated}]{%
    \includegraphics[width=\columnwidth]{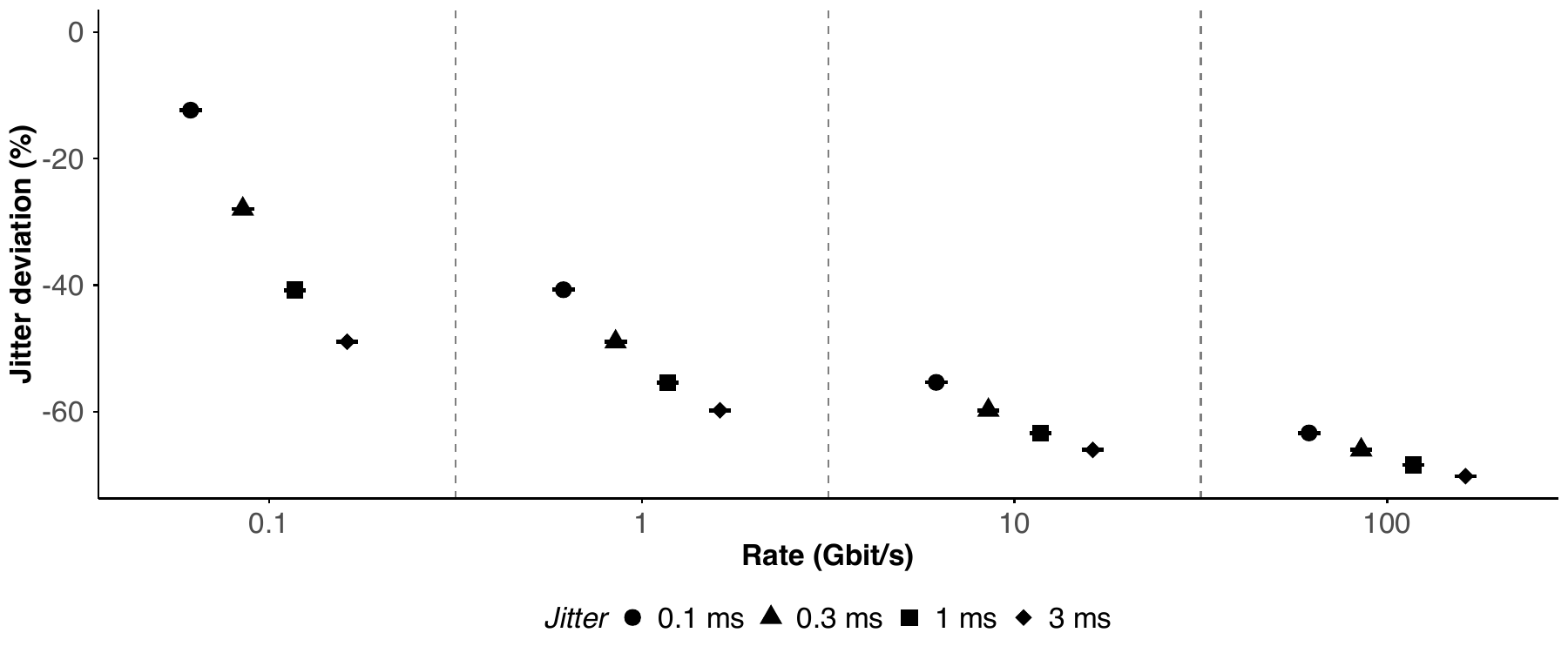}%
  }
  \caption{Simulated accuracy of uncorrelated, non-reordered varying delays for a desired mean delay of \qty{10}{\ms} across various rates and jitters.}
  \label{fig:combined_normal_uncorrelated}
\end{figure}

\autoref{fig:combined_normal_uncorrelated} shows the relative deviations for uncorrelated delays.
The deviation is given in percent, i.e., relative to the desired \qty{10}{\ms} delay or the different jitters.
\autoref{fig:combined_delay_normal_uncorrelated} shows that the achieved mean delay is always larger than desired.
In contrast, \autoref{fig:combined_jitter_normal_uncorrelated} shows that the achieved jitter is always smaller than desired.
The largest deviations exceed $110\%$ for the achieved mean delay and reach $-70\%$ for the achieved jitter.
Both the achieved mean delay and jitter deviate substantially even for a small jitter (\qty{0.1}{\ms}) and traffic rate (\qty{100}{\mbps}).
The deviations worsen for increasing jitter and traffic rate.
Uncorrelated delays are therefore unsuitable for emulating varying delays even for small jitter and traffic rate.

\subsection{Impact of Half-Life Period}
\label{subsec:suitableHalfLife}

We study the impact of the half-life period on delay dynamics and the accuracy of the desired mean and jitter.

\subsubsection{Impact on Delay Dynamics}
\label{subsubsec:delayDynamics}

\begin{figure}[t]
  \centering
  \includegraphics[width=\columnwidth]{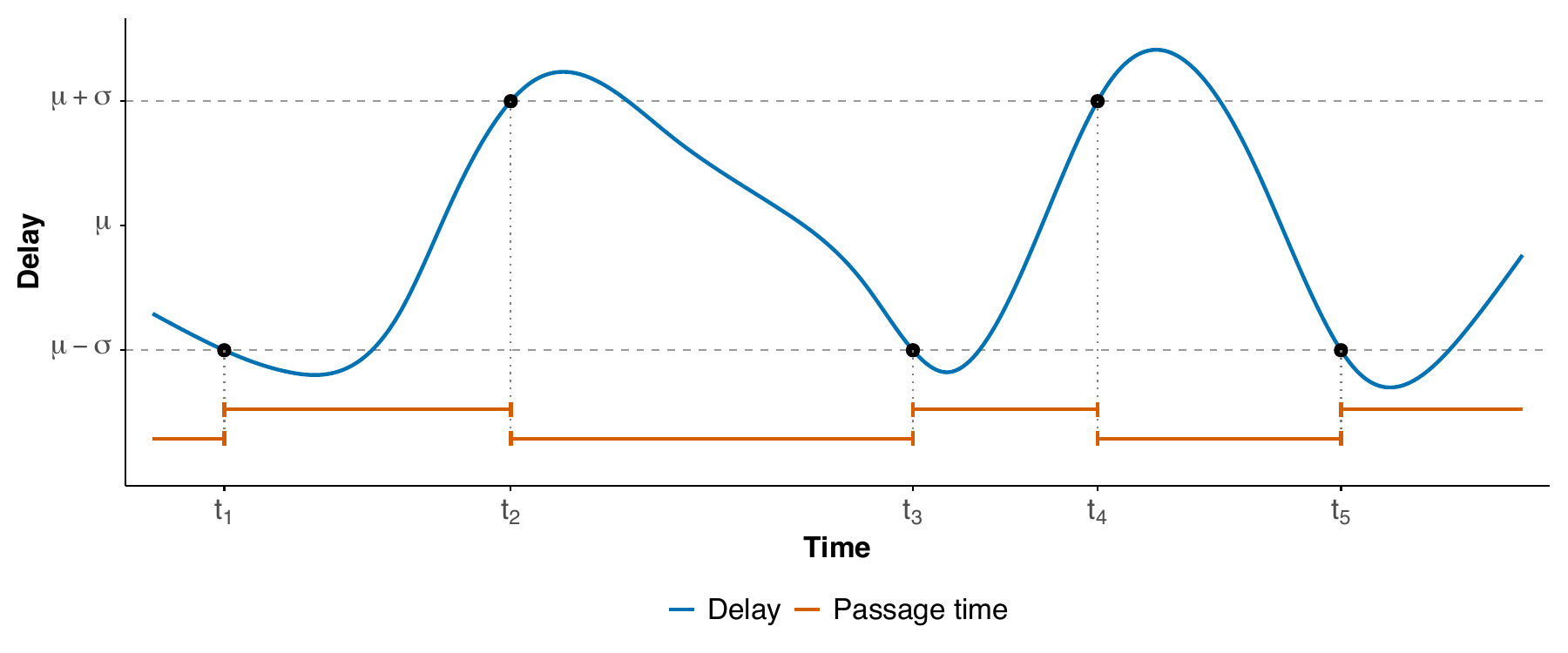}
  \caption{Definition of passage times.}
  \label{fig:passage_time}
\end{figure}

We study how fast a time series of correlated delays changes over time.
To that end, we quantify it by measuring the passage times in these time series.
The concept of passage times is illustrated in \autoref{fig:passage_time}.
A passage time is the duration between the time series exceeding an upper threshold and subsequently falling below a lower threshold, and vice versa.
In our experiments we set the thresholds to $\mu - \sigma$ and $\mu + \sigma$.

\begin{figure}[t]
  \centering
  \includegraphics[width=\columnwidth]{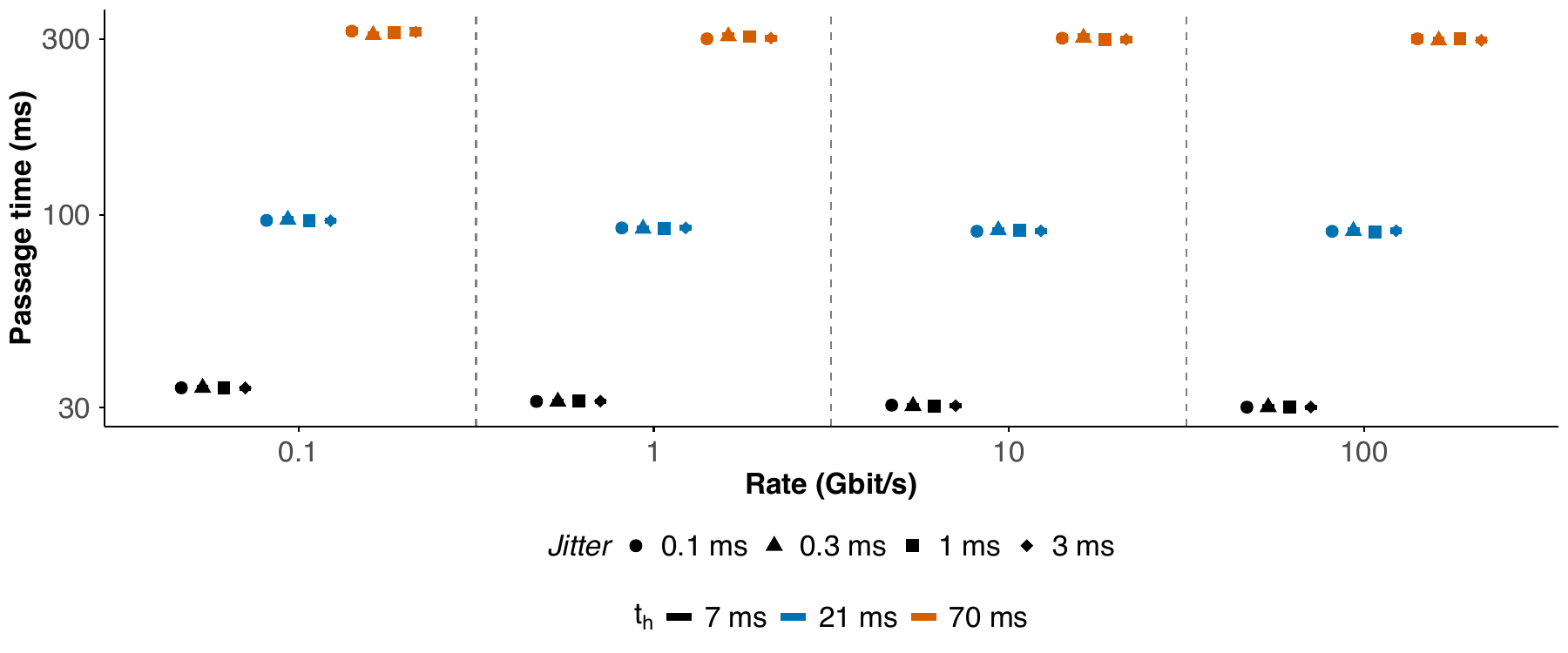}
  \caption{Simulated passage times across various half-life periods, rates, and jitters at a desired mean delay of \qty{10}{\ms}.}
  \label{fig:half_life_normal_passage_time}
\end{figure}

\autoref{fig:half_life_normal_passage_time} shows that the passage time depends only on the half-life period but not on the traffic rate, which was a design goal.
The fact that the passage times are also independent of the desired jitter is caused by the thresholds, which scale with the jitter in our experiments.
The result means that the amplitude of the oscillations of the correlated delays increases linearly with the desired jitter, but their frequency remains the same.
Thus, the dynamics of correlated delay can be well controlled by the half-life period $t_h$.

\subsubsection{Impact on Accuracy}

\begin{figure}[t]
  \centering
  \subfloat[Mean delay accuracy.\label{fig:half_life_normal_delay}]{%
    \includegraphics[width=\columnwidth]{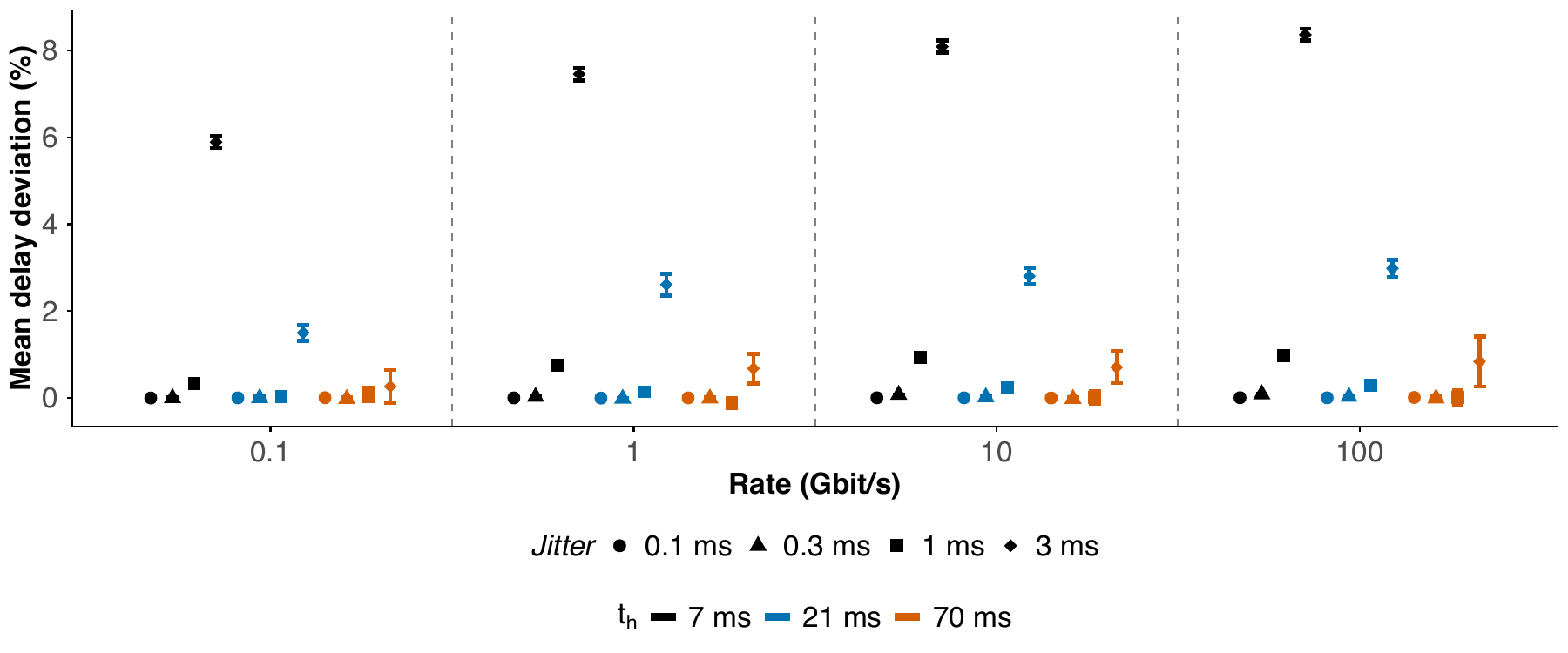}%
  }\\
  \subfloat[Jitter accuracy.\label{fig:half_life_normal_jitter}]{%
    \includegraphics[width=\columnwidth]{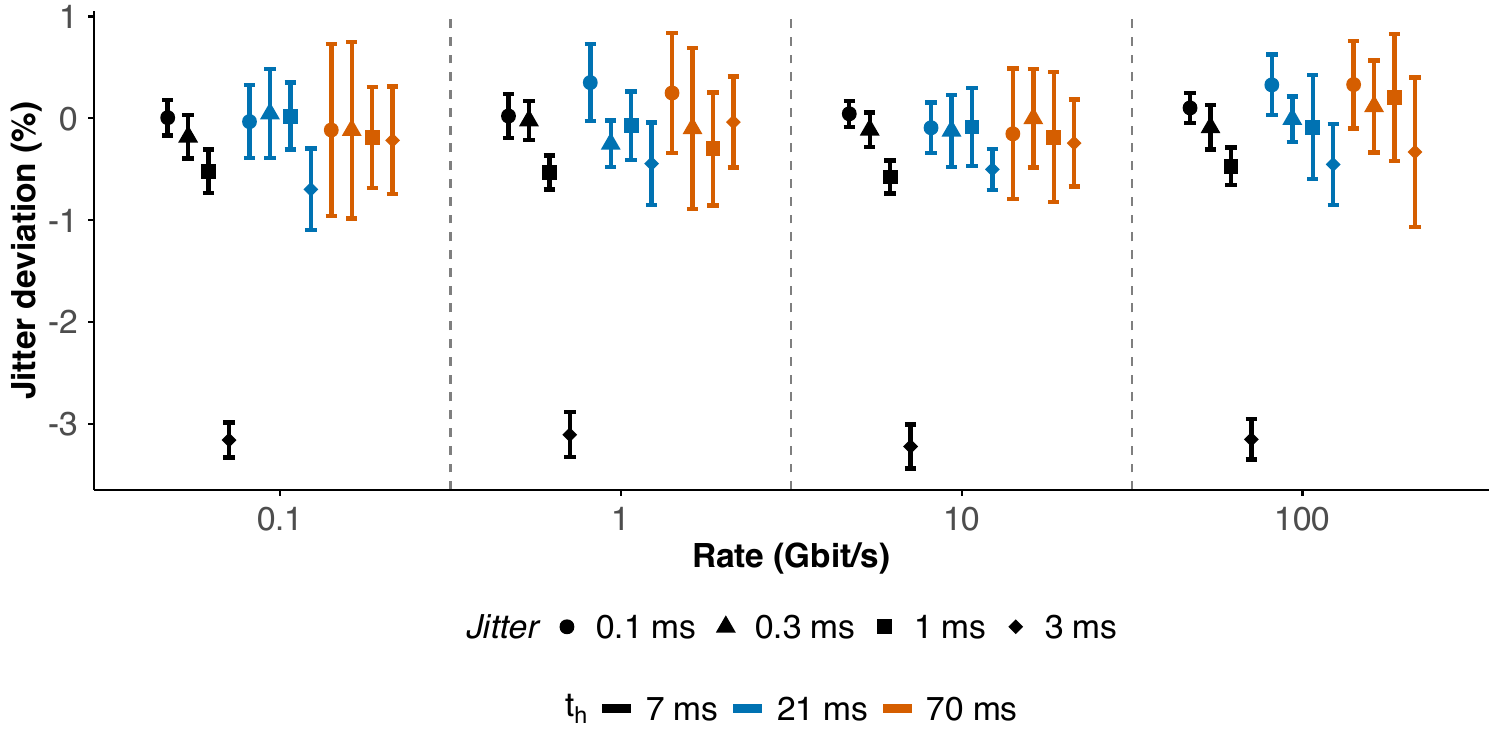}%
  }
  \caption{Simulated accuracy of adaptive delay correlation across various half-life periods, rates, and jitters at a desired mean delay of \qty{10}{\ms}.}
  \label{fig:plot_half_life}
\end{figure}

As outlined in \autoref{subsec:problemVaryingPacketDelay}, accuracy of the achieved packet delay is a major problem when varying packet delay is desired.
Delay correlation significantly mitigates this problem but cannot avoid it.
We investigate the influence of the half-life period and desired jitter on the accuracy of the achieved delay's mean and jitter.

\autoref{fig:half_life_normal_delay} shows the relative deviations of the achieved mean delay, which are all positive, i.e., the achieved delay is only increased compared to the desired mean delay.
We observe that the deviation is largely independent of the traffic rate.
Overall, the deviations are very small, mostly less than 1\%, but a short half-life period (\qty{7}{\ms}) and large jitter (\qty{3}{\ms}) can lead to deviations of 10\% (\qty{1}{\ms}).
While a half-life period of \qty{70}{\ms} minimizes potential inaccuracies, it leads to slower dynamics (see \autoref{fig:half_life_normal_passage_time}).
Thus, we recommend a half-life period of $t_h=\qty{21}{\ms}$ as it leads to good accuracy and keeps passage times sufficiently short for any values of desired jitter.

\autoref{fig:half_life_normal_jitter} illustrates the deviations of the achieved jitter.
The deviations tend to be negative, i.e., achieved jitter is slightly smaller than desired.
This is because short delays are likely to be increased by larger delays of preceding packets, leading to smaller jitter than intended.
Overall, the observed jitter deviations are small, mostly around zero.
However, when the desired jitter is large and the half-life period short, deviations of up to $-3\%$ can occur.
This problem is avoided with our chosen default value of $t_h=\qty{21}{\ms}$.


\subsection{Accuracy Independent of Mean Delay}
\label{subsec:accuracyMeanDelays}

The accuracy of adaptive delay correlation is essentially independent of the desired mean delay.
To demonstrate this, we compare mean delays of \qty{3}{\ms}, \qty{10}{\ms}, and \qty{30}{\ms}.

\begin{figure}[t]
  \centering
  \subfloat[Mean delay accuracy.\label{fig:delays_delay_normal}]{%
    \includegraphics[width=\columnwidth]{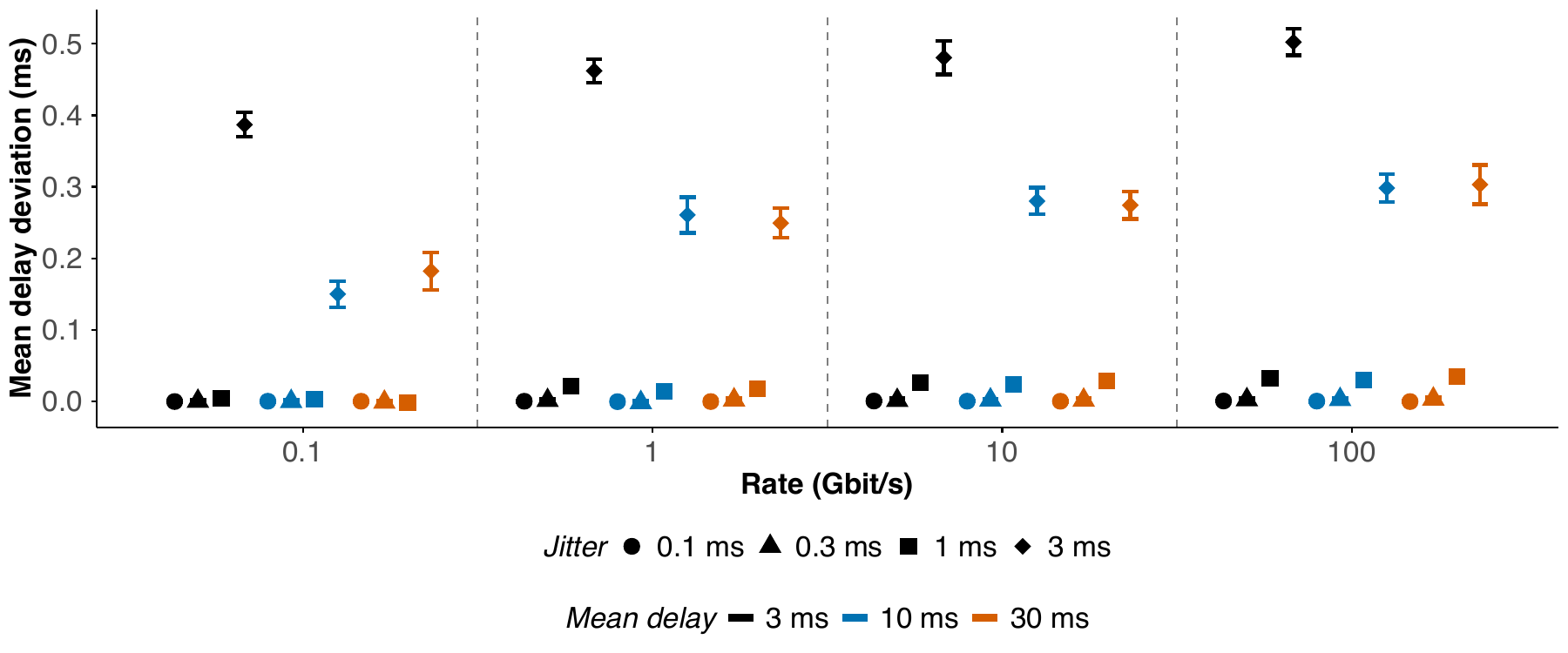}%
  }\\
  \subfloat[Jitter accuracy.\label{fig:delays_jitter_normal}]{%
    \includegraphics[width=\columnwidth]{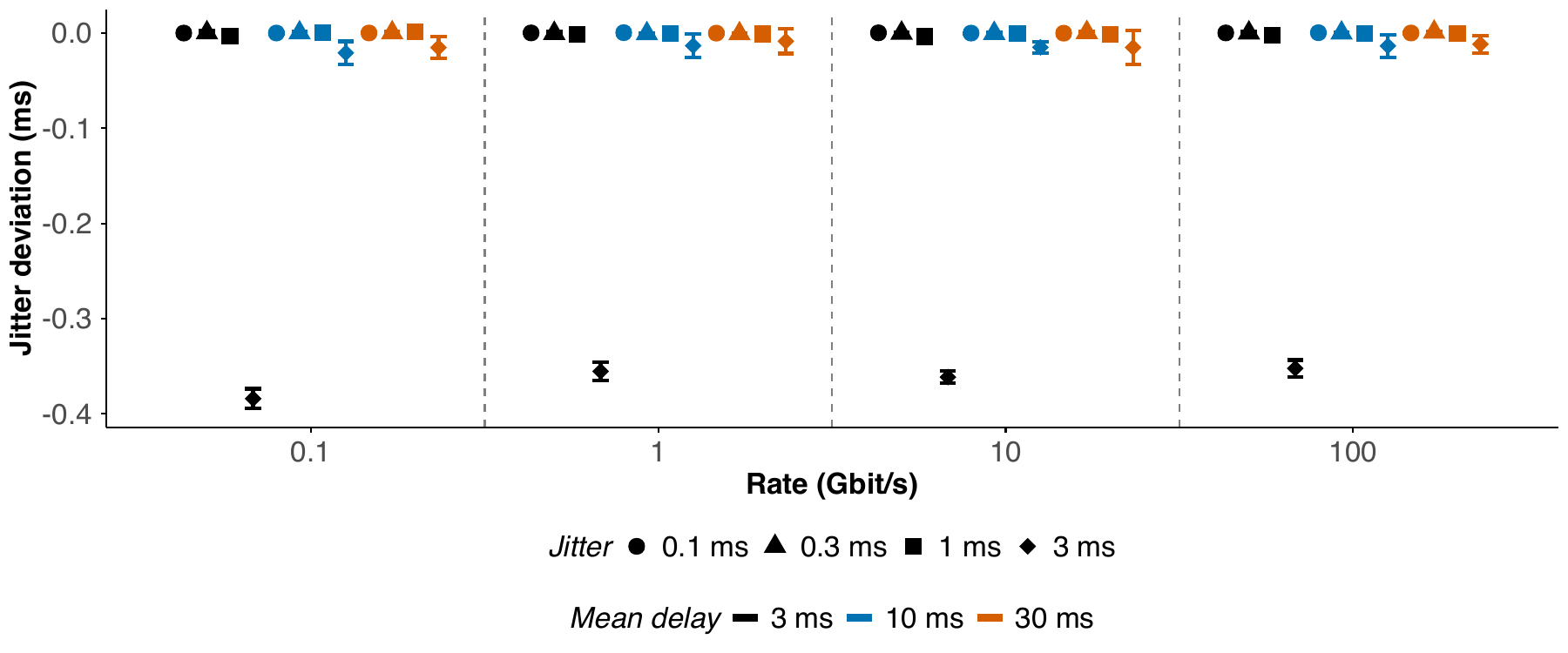}%
  }
  \caption{Simulated accuracy of adaptive delay correlation ($t_h = \qty{21}{\ms}$) across various desired mean delays, rates, and jitters.}
  \label{fig:delays_normal}
\end{figure}

\autoref{fig:delays_normal} shows the absolute deviations (in \unit{\ms}) of the mean delay and jitter for different combinations of the current rate and jitter.
The deviations are largely independent of the desired mean delay and traffic rate.
However, the deviation increases with the desired jitter.
Thus, the results confirm that adaptive correlation with a half-life period of $t_h=\qty{21}{\ms}$ achieves the desired mean delay and jitter.

There is one exception: when the desired jitter approaches the desired mean (\qty{3}{\ms} each).
This problem arises because the normal distribution then frequently generates negative delay values, which cannot be realized in practice.
This increases the achieved mean delay and decreases the achieved delay jitter compared to the desired values.
More complex distributions for delay generation mitigate this problem (see \autoref{subsec:impactDistribution}).


\subsection{Comparison with Packet Reordering}
\label{subsec:comparingAccuracy}

Reordering packets according to their transmission time is NetEm's default configuration.
With packet reordering, packets are sent at intended transmission times and do not have to wait for preceding packets to be sent.
Therefore, the desired mean delay and jitter can be exactly met at the expense of \ac{OOO} delivery.

\begin{figure}[t]
  \centering
  \includegraphics[width=\columnwidth]{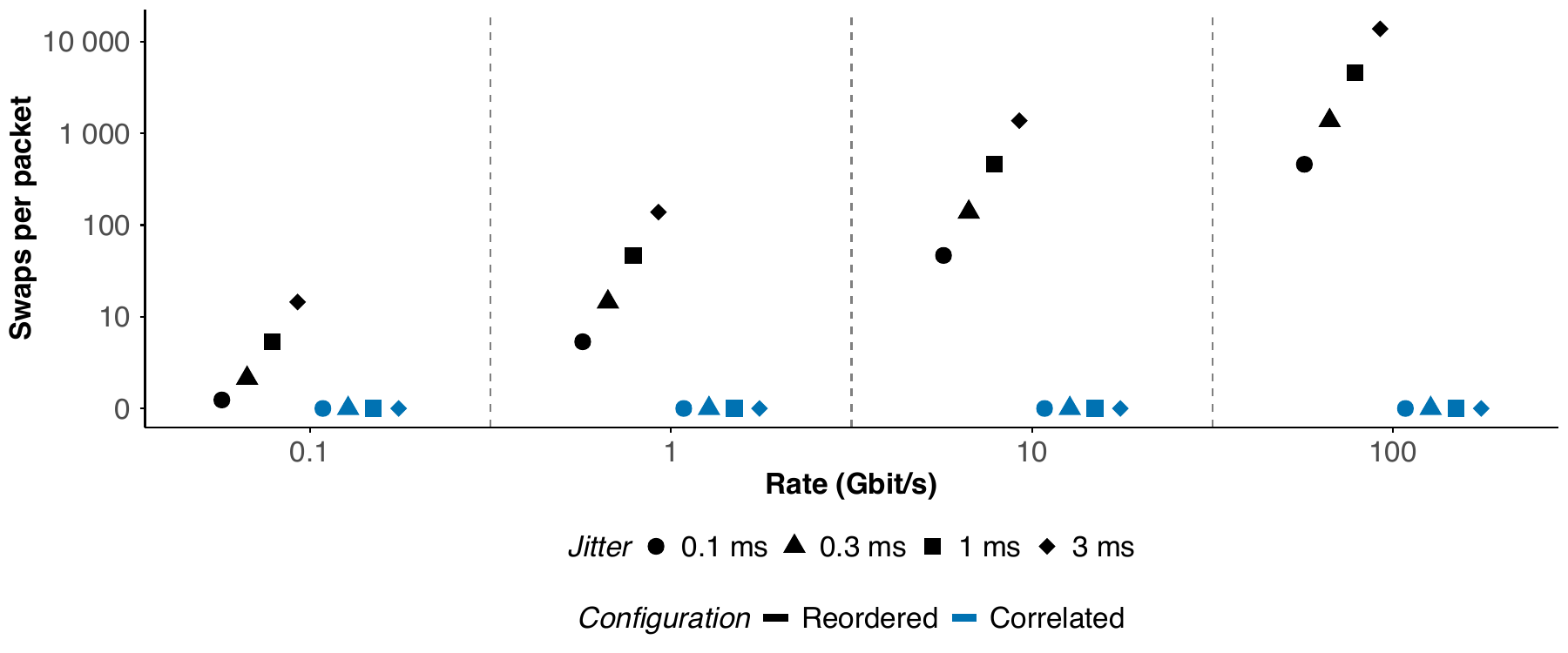}
  \caption{Number of packet swaps required to sort the packets in ascending sequence number order.
    Simulation results for reordering packets according to their transmission time and adaptive delay correlation ($t_h = \qty{21}{\ms}$) at a desired mean delay of \qty{10}{\ms} across various rates and jitters.}
  \label{fig:combined_ktau_normal_reordered_correlated}
\end{figure}

Reordering introduces \ac{OOO} delivery, which negatively impacts the performance of protocols such as TCP~\cite{BlAl02,ZhKa02,LaGe02}, QUIC~\cite{KaJe17}, and RDMA~\cite{MiSh18,SoKh23}.
The severity of this problem depends on how much reordering occurs.
\autoref{fig:combined_ktau_normal_reordered_correlated} shows the number of swaps per packet required to order them into the original sequence.
This is commonly referred to as the Kendall tau or bubble-sort distance (per packet).
Adaptive delay correlation produces no reordering by design, whereas the number of swaps per packet for reordering increases with both the current rate and the desired jitter.
We therefore conclude that reordering should not be the default strategy for varying delay emulation at high traffic rates and jitters.


\subsection{Impact of Limited Bandwidth}
\label{subsec:accuracySpacing}

Limited bandwidth implies positive transmission times and queuing delays for packets, both of which were neglected in the previous experiments.
With limited bandwidth, the achieved packet delay may be increased for both packet reordering and correlated packet delay.
We quantify both in the following.
We consider different utilizations $u\in\{0, 0.5, 0.8, 0.99\}$.
For $u=0$, we reuse the previous values with unlimited bandwidth for comparison.
For $u>0$, the delayed packets are spaced by a virtual queue with unlimited buffer whose processing rate is set to $R/u$, where $R$ is the traffic rate.

\subsubsection{Accuracy of Reordering with Limited Bandwidth}
\label{subsubsec:accuracyPacketReordering}

\begin{figure}[t]
  \centering
  \subfloat[Mean delay accuracy.\label{fig:spacing_delay_reorder_normal}]{%
    \includegraphics[width=\columnwidth]{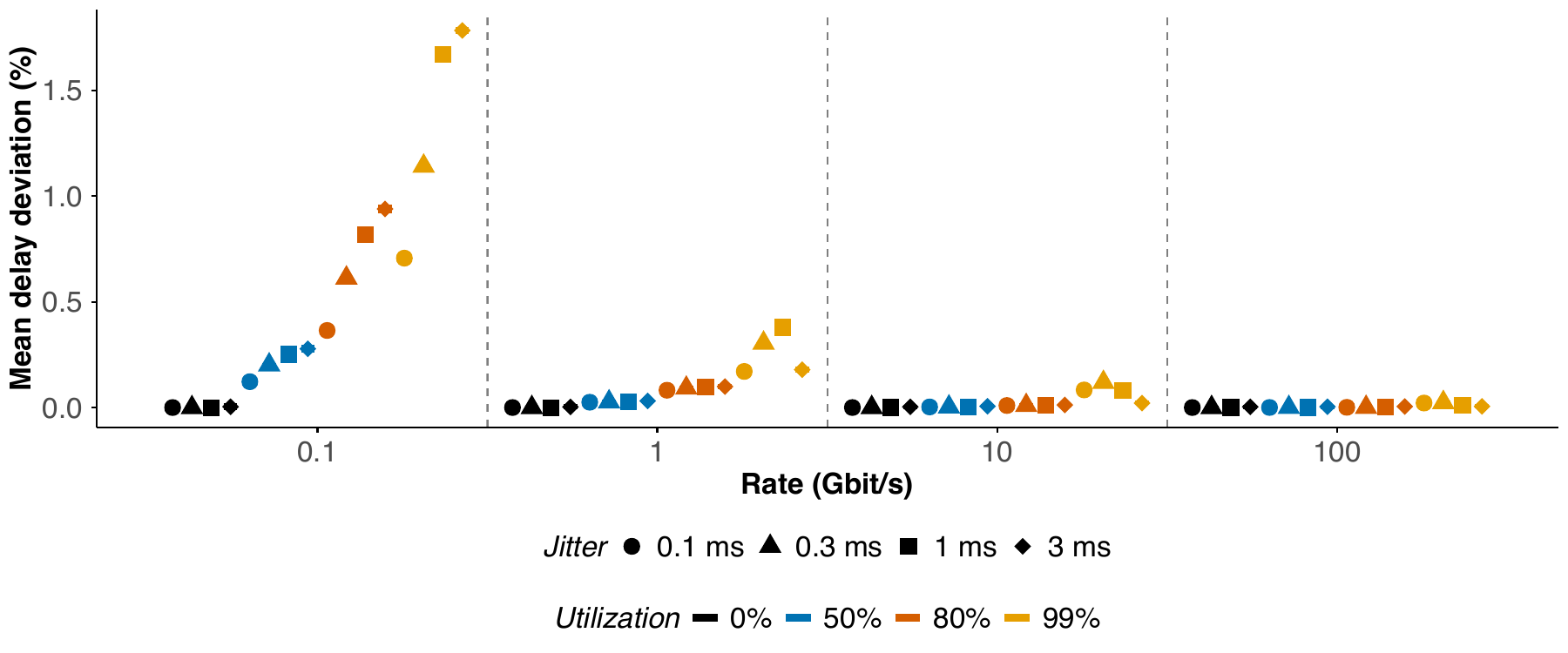}%
  }\\
  \subfloat[Jitter accuracy.\label{fig:spacing_jitter_reorder_normal}]{%
    \includegraphics[width=\columnwidth]{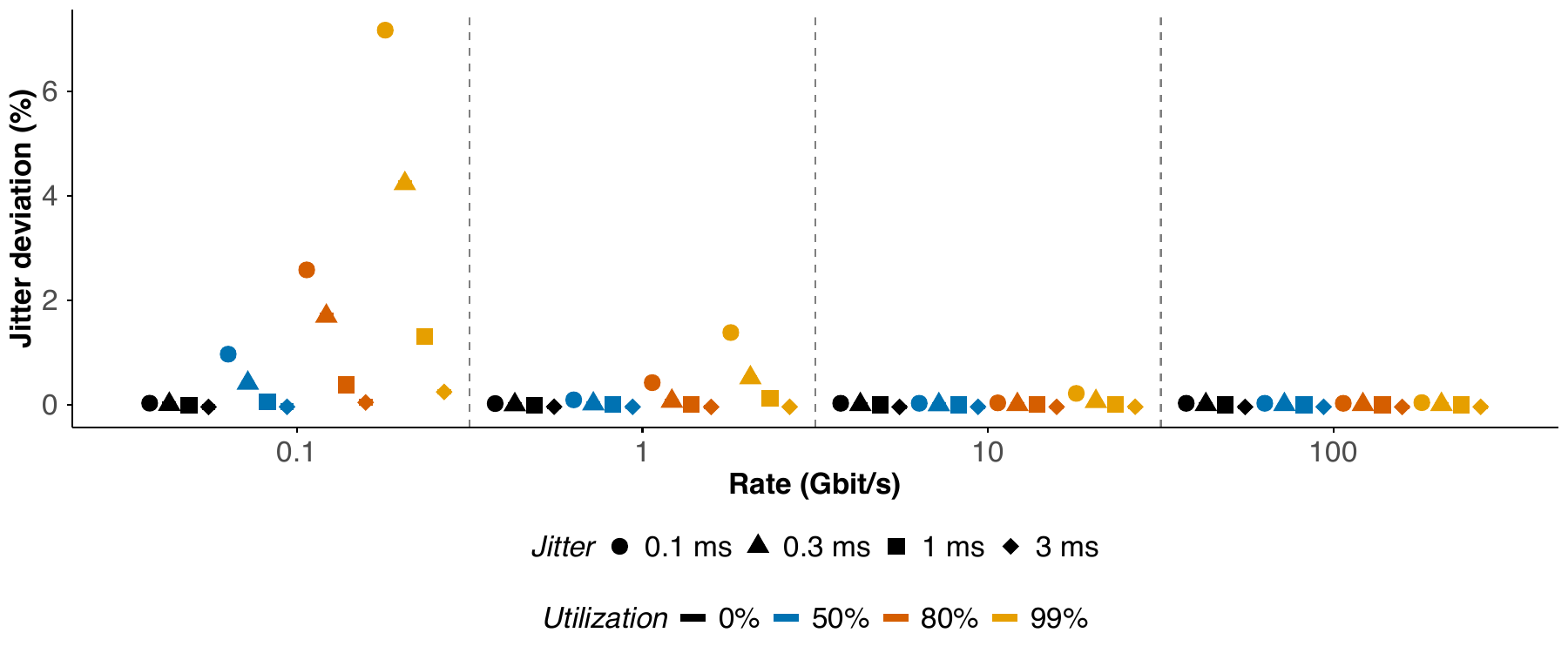}%
  }
  \caption{Simulated accuracy of reordering under limited bandwidth across various utilizations, rates, and jitters for a desired mean delay of \qty{10}{\ms}.}
  \label{fig:spacing_normal_reorder}
\end{figure}

\autoref{fig:spacing_normal_reorder} compiles the relative deviations for mean delay and jitter with packet reordering for different jitters, utilizations, and traffic rates.
We observe that positive mean delay deviations are clearly visible for \qty{100}{\mbps} and \qty{1}{\gbps}, but hardly beyond (see \autoref{fig:spacing_delay_reorder_normal}).
These deviations clearly increase with utilization.
When the packets of a \ac{CBR} stream receive varying delays, the traffic becomes more bursty.
With increasing utilization, the transmission duration of packets increases, which leads to increasing queuing delay for some traffic, causing positive deviations from the desired mean delay.
The deviations of the mean delay also increase with the desired jitter because larger jitter leads to larger bursts.

In contrast, the deviations of the delay jitter decrease with the desired jitter (see \autoref{fig:spacing_jitter_reorder_normal}).
We observe this because many small delays are extended by queuing, which reduces the delay variation.
Further, with packet reordering, the achieved mean delay and jitter may be larger than desired, but only for small traffic rates.
For large traffic rates, the reordered traffic stream is not bursty enough to add substantial queuing delay to the traffic so that the desired mean delay and jitter are again almost perfectly met even after queuing.

\subsubsection{Accuracy of Correlated Delays with Limited Bandwidth}
\label{subsubsec:adaptiveCorrelationWithSpacing}

\begin{figure}[t]
  \centering
  \subfloat[Mean delay accuracy.\label{fig:spacing_delay_correlation30_normal}]{%
    \includegraphics[width=\columnwidth]{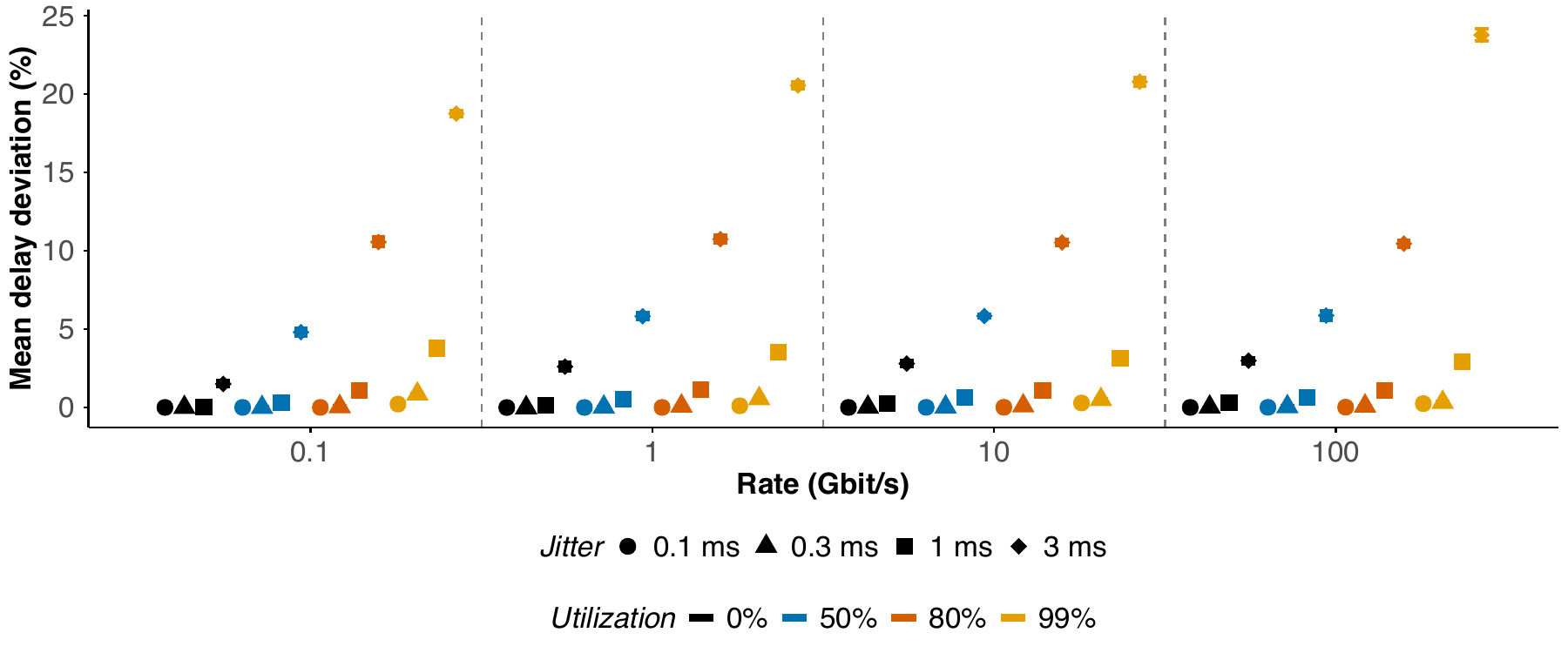}%
  }\\
  \subfloat[Jitter accuracy.\label{fig:spacing_jitter_correlation30_normal}]{%
    \includegraphics[width=\columnwidth]{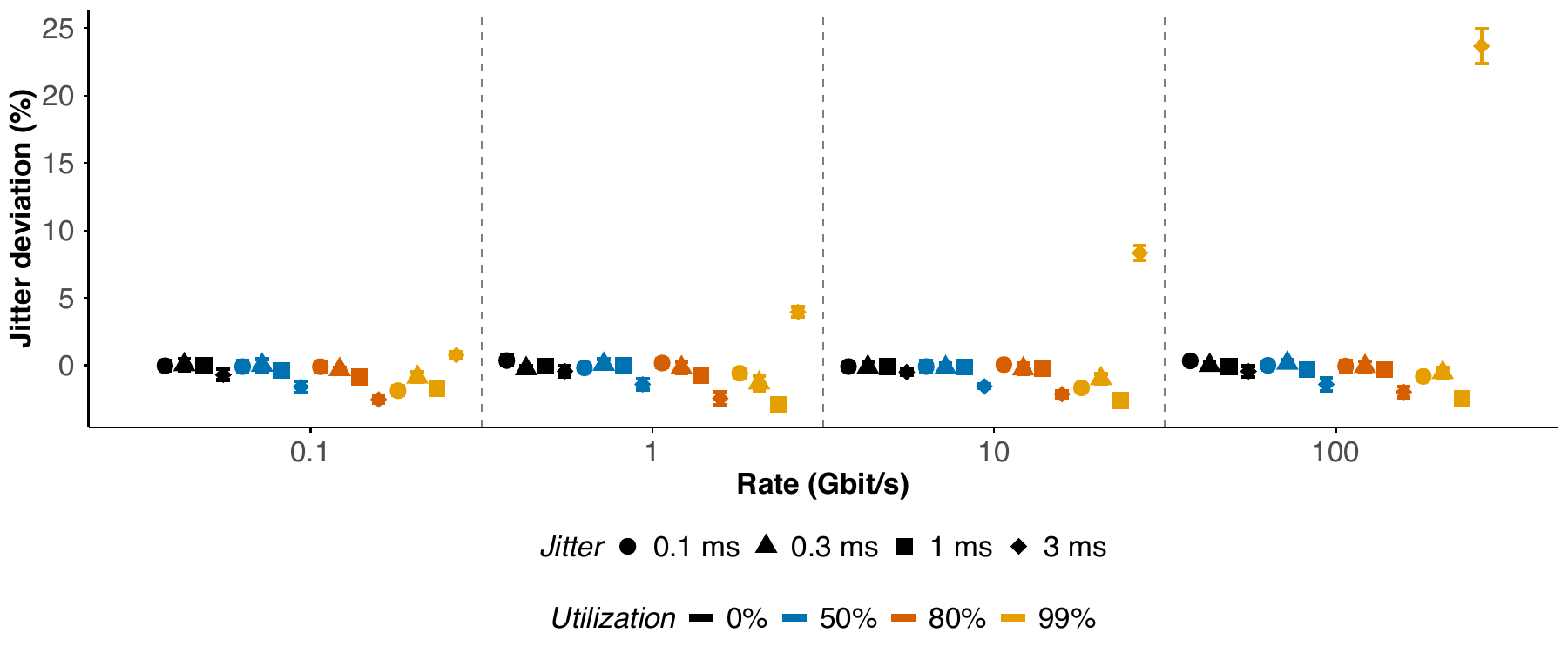}%
  }
  \caption{Simulated accuracy of adaptive delay correlation ($t_h = \qty{21}{\ms}$) under limited bandwidth across various utilizations, rates, and jitters for a desired mean delay of \qty{10}{\ms}.}
  \label{fig:spacing_normal_correlation}
\end{figure}

\autoref{fig:spacing_normal_correlation} shows the relative deviations for mean delay and jitter with correlated packet delays.
The deviations are significantly larger than with packet reordering and do not vanish with increasing traffic rates.
Apart from that, they also increase with utilization and desired jitter.

The reason for the increased mean delay deviation (see \autoref{fig:spacing_delay_correlation30_normal}) is the increased queuing delay.
With packet delaying and reordering, a \ac{CBR} traffic stream is turned into a relatively smooth traffic stream which is fed into the virtual queue so that relatively little queuing delay occurs.
With correlated packet delays and without reordering, significant packet bursts may occur as differently delayed packets need to wait for each other, leading to substantially more queuing delay at the virtual queue.
This effect increases with queue utilization.

The jitter is mostly slightly decreased (see \autoref{fig:spacing_jitter_correlation30_normal}) because small delays are increased by queuing.
Only for very high utilization of $u=99\%$ and large jitter (\qty{3}{\ms}), the achieved jitter is increased due to excessive queuing delays.
In summary, the accuracy of correlated packet delay can be significantly impaired by limited bandwidth, but only if the desired jitter is very large and the utilization is very high.
For all other parameter settings, desired mean delay and jitter can be well met with correlated packet delays.

\subsection{Impact of Delay Distribution}
\label{subsec:impactDistribution}

We study the impact of the underlying delay distribution on the dynamics, accuracy, and sample generation rate of correlated delays with a desired jitter of \qty{3}{\ms}.
The distributions we investigate are the normal, uniform, gamma, and log-normal distribution.

\subsubsection{Impact on Delay Dynamics}
\label{subsubsec:distributionDynamics}

\begin{figure}[t]
  \centering
  \includegraphics[width=\columnwidth]{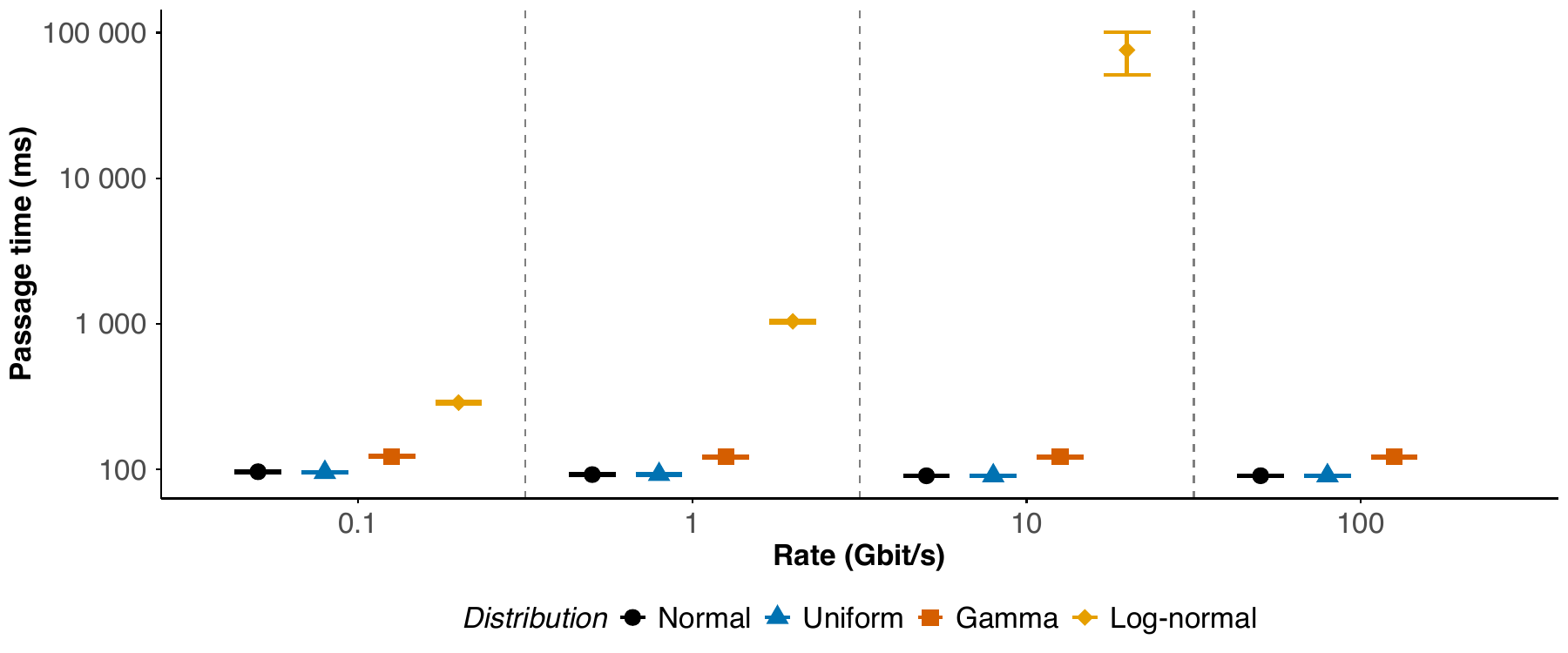}
  \caption{Simulated passage times across various distributions and rates, at a desired mean delay of \qty{10}{\ms} and jitter of \qty{3}{\ms}.}
  \label{fig:distributions_passage_time}
\end{figure}

\autoref{fig:distributions_passage_time} compiles the measured passage times of different distributions.
As shown in \autoref{subsubsec:delayDynamics}, the passage time depends only on the configured half-life period and not on the traffic rate.
The passage times of the normal and uniform distributions are similar, while the gamma distribution has a slightly higher one.

One exception is the log-normal distribution, whose passage time increases with the traffic rate.
For \qty{100}{\gbps}, no passage time is measured during \qty{10}{\min} of simulation.
We observe this because the log-normal distribution has a particularly long tail, especially after the variance scaling applied by adaptive delay correlation.
The very high delays that frequently occur add significant queuing delay to successive packets, and smaller delays cannot be realized in practice.

\subsubsection{Impact on Accuracy}

\begin{figure}[t]
  \centering
  \subfloat[Mean delay accuracy.\label{fig:distributions_delay}]{%
    \includegraphics[width=\columnwidth]{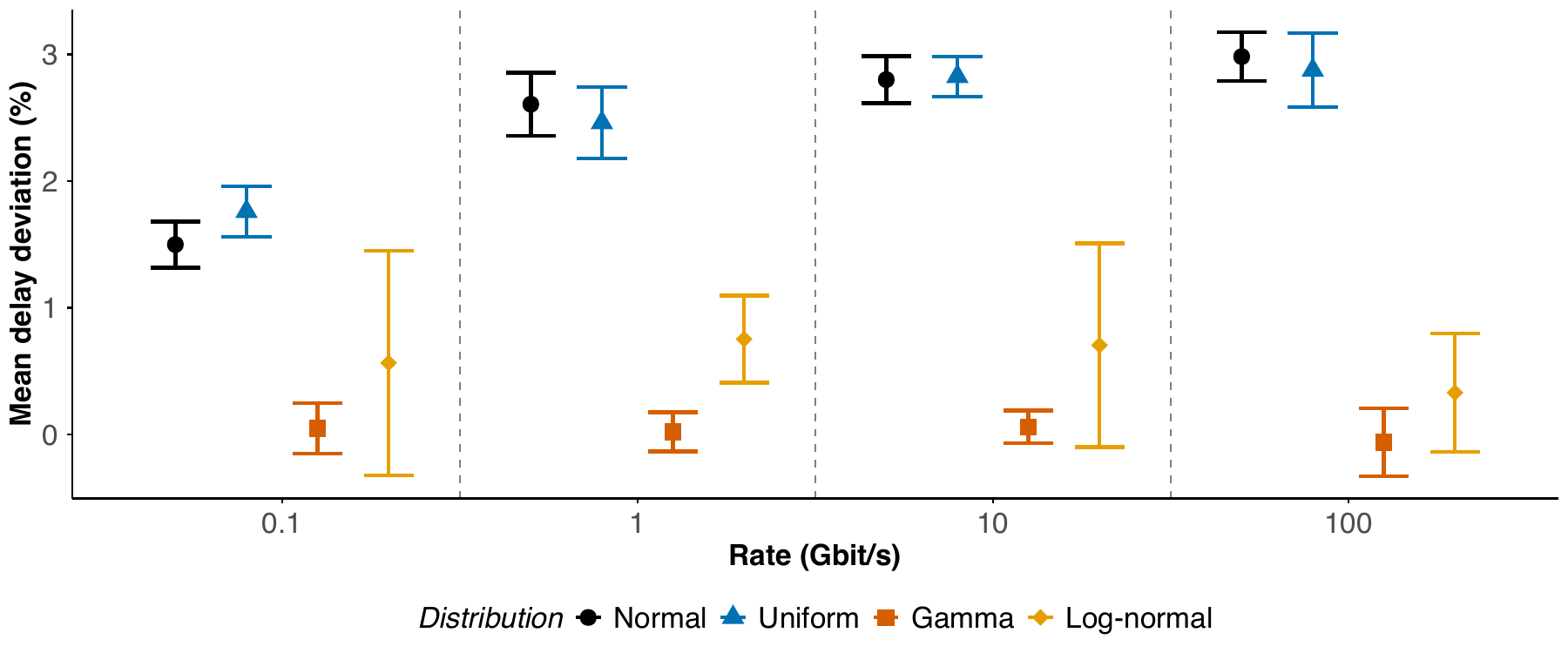}%
  }\\
  \subfloat[Jitter accuracy.\label{fig:distributions_jitter}]{%
    \includegraphics[width=\columnwidth]{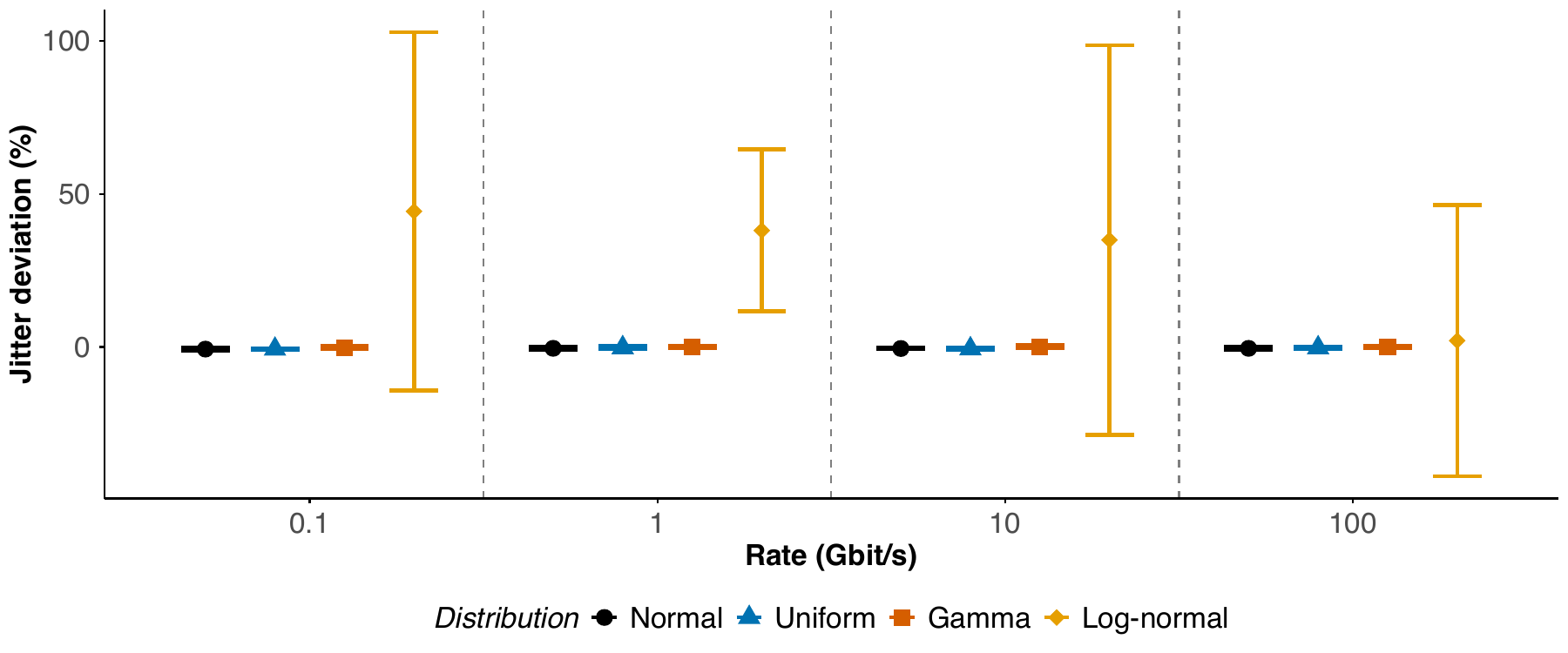}%
  }
  \caption{Simulated accuracy of adaptive delay correlation ($t_h = \qty{21}{\ms}$) under various delay distributions and rates, for a desired mean delay of \qty{10}{\ms} and jitter of \qty{3}{\ms}.}
  \label{fig:distributions_accuracy}
\end{figure}

The shape of the underlying delay distribution impacts the accuracy of the achieved mean delay and jitter.
\autoref{fig:distributions_accuracy} compiles the relative deviations of the mean delay and jitter for the different distributions.
The mean delay deviations of all four distributions are within the same range, while the jitter deviation of the log-normal distribution is much larger than those of the other distributions.

\autoref{fig:distributions_delay} shows that the normal and uniform distributions both achieve a slightly larger mean delay.
We observe this because both frequently generate negative delays that cannot be realized in practice (see \autoref{subsec:accuracyMeanDelays}).
In contrast, the gamma distribution nearly perfectly emulates the desired mean delay with less than 0.1\% deviation.
The normal, uniform, and gamma distributions achieve a jitter deviation of less than 1\% (see \autoref{fig:distributions_jitter}).
Therefore, the gamma distribution further increases the accuracy of adaptive delay correlation.

We do not recommend using the log-normal distribution for adaptive delay correlation.
Its realized mean delay is more accurate than those of the normal and uniform distributions (see \autoref{fig:distributions_delay}).
However, its realized jitter deviation significantly exceeds those of the other distributions.
Further, its large passage times (see \autoref{subsubsec:distributionDynamics}) result in a realized mean delay and jitter that are not converged after \qty{10}{\min}.
Therefore, the long tail of the log-normal distribution makes it unsuitable for adaptive delay correlation.

\subsubsection{Impact on Sample Generation Rate}

\begin{figure}[t]
  \centering
  \includegraphics[width=0.7\columnwidth]{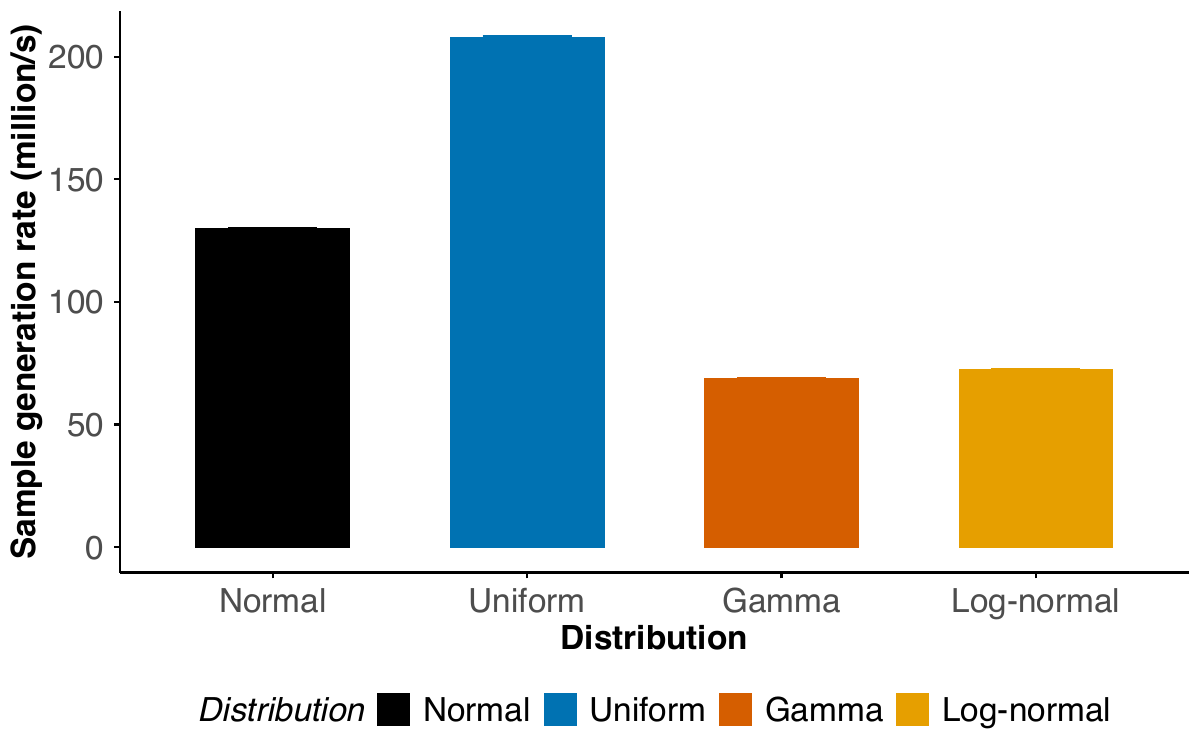}
  \caption{Simulated sampling rate across various distributions at a desired mean delay of \qty{10}{\ms} and jitter of \qty{3}{\ms}.}
  \label{fig:distributions_sampling_cost}
\end{figure}

Generating each delay sample takes some time, which was neglected in the previous experiments.
To evaluate this generation time, we generate one billion delay samples and measure the time it takes.
The samples are generated on the same server used for the prototype evaluation (see \autoref{subsubsec:testbedConfiguration}).

\autoref{fig:distributions_sampling_cost} compiles the average sample generation rate in million per second.
The normal distribution's generation rate serves as a baseline since most \acp{NLE} use normally distributed delays.
Uniformly distributed samples are easier to generate, resulting in a generation rate that is $1.5\times$ that of normally distributed ones.
In contrast, generating samples from a gamma or log-normal distribution is more complex than from a normal distribution, halving the generation rates.
Despite its accuracy advantage, the gamma distribution is not chosen as the default, because this advantage does not justify the resulting generation overhead.

%% file: content/dpds.tex
\section{The DPDK-Based Packet Delayer and Spacer (DPDS)}
\label{sec:theDPDSEmulator}

We developed the \acf{DPDS} for link emulation with a kernel bypass approach.
\ac{DPDS} emulates constant and varying delays, bandwidth limitation, and packet loss.
It applies those characteristics to a single link regardless of the number of flows.
\ac{DPDS} is open source on GitHub~\cite{GitHubDPDS}.

\subsection{General Architecture}

\begin{figure}[t]
  \centering
  \includegraphics[width=\columnwidth]{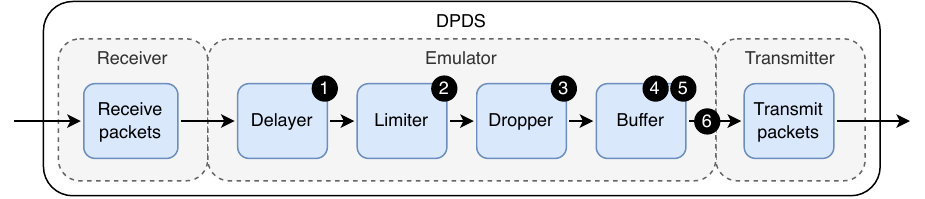}
  \caption{
    General architecture of \acs{DPDS}, consisting of receiver, emulator, and transmitter components.
  }
  \label{fig:generalConceptDPDS}
\end{figure}

\ac{DPDS} is organized as a three-stage pipeline consisting of a receiver, emulator, and transmitter stage (see \autoref{fig:generalConceptDPDS}).
The receiver and transmitter stages implement the packet reception and transmission, respectively.
The emulator stage implements all supported emulation features, such as packet delay, limited bandwidth, and packet loss.

When packets enter \ac{DPDS}'s emulator stage, they are processed in six steps (see \autoref{fig:generalConceptDPDS}):
\begin{itemize}
  \item[\circled{1}] The delays with or without adaptive delay correlation are calculated (see \autoref{subsubsec:adaptiveCorrelation}).
  \item[\circled{2}] The packets are either policed (see \autoref{subsubsec:policing}) or spaced (see \autoref{subsubsec:spacing}) to enforce a bandwidth limit.
  \item[\circled{3}] Packets are dropped according to the configured packet loss model (see \autoref{subsec:packetLoss}).
  \item[\circled{4}] The packets' transmission times are calculated by adding the delay and the spacing offset to the current time.
  \item[\circled{5}] The packets and their respective transmission times are stored in a buffer.
  \item[\circled{6}] Packets are sent once the current time exceeds their transmission time.
\end{itemize}

\ac{DPDS} can emulate link characteristics on a single port or between two separate ports of a host.
It uses software timestamps, eliminating the need for special \ac{NIC} features.
The accuracy of software timestamps is evaluated in \autoref{subsec:accuracyVaryingDelay}.
To enable high emulation rates, \ac{DPDS} processes packets in batches of up to 16, which increases throughput at the cost of reduced delay accuracy and increased burstiness.
Additionally, the receiver, emulator, and transmitter (\autoref{fig:generalConceptDPDS}) can run as a single thread or as up to three threads on separate CPU cores.

\subsection{Packet Delays}
\label{subsec:packetDelays}

\ac{DPDS} implements packet delays by using a parameterized distribution combined with either reordering or adaptive delay correlation.
In the following, we describe which delay distributions \ac{DPDS} supports.
Then, we explain how \ac{DPDS} implements reordering and adaptive delay correlation.

\subsubsection{Supported Delay Distributions}

\ac{DPDS} supports emulating both constant and varying delays.
Constant delays are implemented using a deterministic distribution.
Varying delays are implemented using one of the following distributions: uniform, normal, log-normal, and gamma.
These distributions are parameterized using a mean delay and jitter.
The uniform and normal distributions are supported to match those of existing \acp{NLE}.
The gamma distribution is supported since Mukherjee~\cite{Mu94} showed that Internet delays follow a shifted gamma distribution.

\subsubsection{Reordering}

\ac{DPDS} supports packet reordering by sorting packets by their transmission time.
This is implemented using a binary heap for the packet buffer, which automatically sorts the packets as they are pushed into the data structure.
While this enables accurate emulation of varying delays, it also introduces significant sorting overhead.
This overhead makes reordering infeasible for high rates (see \autoref{subsec:maximumSupportedRates}).
Furthermore, it introduces \ac{OOO} delivery, making it unsuitable for protocols like TCP, QUIC, and RDMA.
If reordering is disabled, the packet buffer is implemented as a simple \ac{FIFO} queue.

\subsubsection{Adaptive Delay Correlation}
\label{subsubsec:adaptiveCorrelation}

\ac{DPDS} implements adaptive delay correlation as described in \autoref{sec:adaptiveCorrelatedPacketDelays}.
To this end, it correlates successive delays through an \ac{EMA} whose weight is recomputed periodically from a short-window rate estimate.
Unlike reordering, this preserves packet order and avoids sorting overhead, enabling higher forwarding rates (see \autoref{subsec:maximumSupportedRates}).
Adaptive delay correlation is parameterized by the mean delay $\mu$ and jitter $\sigma$ of the underlying distribution, as well as the half-life period $t_h$.
The resulting accuracy of \ac{DPDS} is evaluated under both constant (\autoref{subsec:accuracyVaryingDelay}) and changing (\autoref{subsec:measuredAccuracyChangingRate}) rates.

\subsection{Bandwidth Limitation}

\ac{DPDS} implements bandwidth limitation by either policing or spacing packets.
With policing, packets that exceed the configured rate are dropped.
In contrast, spacing delays packets to meet the configured rate.

\subsubsection{Policing}
\label{subsubsec:policing}

Policing is implemented using the token bucket algorithm~\cite{RFC2698}.
In this algorithm, each bit corresponds to a token.
The bucket fill state increases at the configured token rate in proportion to the time elapsed since the last increase.
However, the fill state cannot exceed a configured threshold.
When a packet is processed and there are enough tokens available, the fill state decreases according to the packet's length in bits (tokens).
If a packet's length exceeds the fill state, the packet is dropped.

\subsubsection{Spacing}
\label{subsubsec:spacing}

Spacing is implemented using the leaky bucket algorithm~\cite{Tu86}.
In contrast to the token bucket algorithm, the bucket fill state represents the number of packets currently scheduled for transmission rather than the available tokens, i.e., the bandwidth budget.
Each packet arrives at the spacer with a transmission time determined by the delayer (see \autoref{subsec:packetDelays}).
The spacer then adds an additional offset to this time so that the resulting transmission time adheres to the configured rate.
This offset is derived from the current fill state and the configured rate.
To prevent unbounded delays, the fill state is subject to a configurable upper limit.
Packets that would exceed this limit are dropped.

\subsection{Packet Loss}
\label{subsec:packetLoss}

\ac{DPDS} currently supports two different loss models: independent packet losses and the Gilbert-Elliott model~\cite{Gi60,El63,HaHo08}.
For independent losses, each packet has an independent probability $p$ of being lost.
In contrast, the Gilbert-Elliott model uses a two-state Markov chain, with state-dependent loss probabilities, to emulate correlated packet losses.

%% file: content/evaluation.tex
\section{Evaluation of the DPDS Prototype}
\label{sec:prototypeEvaluation}

We experimentally evaluate \ac{DPDS} along two dimensions.
First, we measure \ac{DPDS}'s maximum forwarding rate and \ac{ZLT} and compare them with NetEm and MoonEm.
Those serve as baselines for in-kernel and kernel bypass-based emulators, respectively.
Second, we evaluate \ac{DPDS}'s accuracy and compare the results with our simulation predictions, including changing traffic rates.

\subsection{Methodology for Measuring Emulator Accuracy}
\label{subsec:methodologyForMeasuringEmulatorAccuracy}

In the following, we describe our experimental setup to facilitate reproducibility, consisting of the traffic generator P4TG, the measurement setup, and the testbed configuration.

\subsubsection{The Traffic Generator P4TG}

P4TG~\cite{LiHa23} is a P4-based, hardware-accelerated traffic generator implemented on the Intel Tofino\texttrademark switching ASIC.
It supports aggregate generation rates of up to \qty{4}{\tbps}~\cite{IhZi25}.
It performs \ac{RTT} measurements with nanosecond granularity directly in the data plane, enabling precise calculation of the mean, standard deviation, and percentiles without sampling bias~\cite{IhZi25_2}.
P4TG also supports automated \ac{ZLT} determination~\cite{IhZi26}.
The \ac{ZLT} is the maximum rate at which a device under test forwards packets without loss, as defined in RFC 2544~\cite{RFC2544}.
P4TG's source code is available on GitHub~\cite{GitHubP4TG}.

\subsubsection{Measurement Setup}

\begin{figure}[t]
  \centering
  \includegraphics[width=.8\columnwidth]{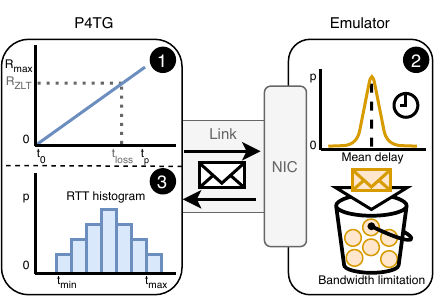}
  \caption{Setup for measuring an emulator's accuracy.}
  \label{fig:measurementSetup}
\end{figure}

\autoref{fig:measurementSetup} depicts the general measurement setup which can be used to evaluate an emulator's accuracy.
P4TG is directly connected to the emulator under test with a single link.
In the first step~\circled{1}, P4TG measures the \ac{ZLT} and maximum forwarding rate of the emulators for the evaluated configuration.
In contrast to the \ac{ZLT}, which is the highest rate at which no packets are lost, the maximum forwarding rate is the highest output rate an emulator sustains under load.
In the second step \circled{2}, P4TG generates traffic and forwards it to the emulator under test.
Based on the rates of the first step, we ensure not to overload the emulator.
The emulator applies the desired link characteristics, such as packet delay and bandwidth limitation.
After applying these characteristics, the emulator forwards the traffic back to P4TG.
In the third step \circled{3}, P4TG uses the \ac{RTT} histogram to measure the emulated mean delay and jitter.
This histogram is then exported via P4TG's REST API for further use.

\subsubsection{Testbed Specifications and Configuration}
\label{subsubsec:testbedConfiguration}

We evaluate \ac{DPDS} using the measurement setup described above.
P4TG \texttt{v2.7.1} runs on an Intel(R) Tofino 1 and generates the traffic.
\ac{DPDS} runs on a bare-metal server with an eight-core Intel(R) Xeon(R) Gold 6134 CPU @ \qty{3.20}{GHz}, four $\qty{32}{\gB}$ DIMM DDR4 RAM modules, and one NUMA node.
The server is equipped with one dual-port (\qty{100}{\gbps} per port) ConnectX-5 MCX516A-CCAT \ac{NIC} with PCIe Gen4 and the MLNX OFED \texttt{23.10} driver.
For efficient packet processing, the server uses $64\times\qty{1}{\gB}$ huge pages, isolated CPU cores, and a PCIe maximum read request size of \qty{1024}{B}.
Its operating system is Ubuntu \texttt{24.04.4 LTS} with kernel \texttt{6.8.0-100-generic}, and we use DPDK \texttt{v25.11} and Rust \texttt{v1.95}.

We configure all emulators under test with a mean delay of \qty{10}{\ms} and a normal and deterministic distribution for varying and constant delay, respectively.
For \ac{DPDS}, when spacing is active, we set the spacing rate to \qty{100}{\gbps}, matching the maximum supported rate of the \ac{NIC}.
\ac{DPDS} and NetEm run on a single, isolated CPU core, whereas MoonEm requires two CPU cores by design.
We run each experiment with 1518-byte packets for \qty{4}{\min} and repeat it ten times to compute confidence intervals with a confidence level of 95\%.


\subsection{Maximum Supported Rates}
\label{subsec:maximumSupportedRates}

We compare the maximum forwarding rate and \ac{ZLT} of \ac{DPDS} with those of NetEm and MoonEm.
NetEm serves as a baseline for in-kernel Linux \ac{QDisc} emulators, and MoonEm serves as a baseline for kernel bypass-based emulators.
Other emulators in each class (e.g., 6GDetCom\_Emulator, TheaterQ, DEMU, SmartNet) may differ in performance due to architectural choices such as SmartNIC offloading.
However, benchmarking them is beyond the scope of this evaluation.

\subsubsection{Forwarding Rate}

In the following, we compare the measured maximum forwarding rates of \ac{DPDS} with those of NetEm and MoonEm.
\autoref{tab:maxForwardingRates} presents the measured maximum forwarding rates for jitters of \qty{0}{\ms} and \qty{3}{\ms}.

\renewcommand{\cellalign}{cc}
\begin{table}[t]
  \caption{Measured maximum forwarding rates of NetEm, MoonEm, and DPDS.}
  \label{tab:maxForwardingRates}
  \centering
  \begin{tabularx}{\columnwidth}{l|YY}
    \toprule
                          & \multicolumn{2}{c}{\emph{Jitter}}                       \\
                          & \emph{\qty{0}{\ms}}               & \emph{\qty{3}{\ms}} \\
    \midrule
    NetEm                 & \qty{5.1}{\gbps}                  & \qty{4.7}{\gbps}    \\
    NetEm + TBF           & \qty{4.7}{\gbps}                  & \qty{4.3}{\gbps}    \\
    MoonEm                & \qty{46.0}{\gbps}                 & $\times$            \\
    DPDS (reordered)      & \qty{78.5}{\gbps}                 & \qty{75.8}{\gbps}   \\
    DPDS (corr.)          & \qty{95.5}{\gbps}                 & \qty{93.6}{\gbps}   \\
    DPDS (corr. + spaced) & \qty{94.8}{\gbps}                 & \qty{94.9}{\gbps}   \\
    \bottomrule
  \end{tabularx}
\end{table}

\paragraph{NetEm}

NetEm's maximum forwarding rate is \qty{5.1}{\gbps} and \qty{4.7}{\gbps} for \qty{0}{\ms} and \qty{3}{\ms} jitter, respectively.
The lower rate at \qty{3}{\ms} jitter is due to the additional sorting overhead that NetEm incurs for varying delays.
A further decrease occurs when NetEm is combined with \ac{TBF} at a spacing rate of \qty{10}{\gbps}: the rates drop to \qty{4.7}{\gbps} and \qty{4.3}{\gbps} for \qty{0}{\ms} and \qty{3}{\ms} jitter, respectively.
This decrease is due to the additional per-packet processing overhead from calculating the spacing offset and applying it.

\paragraph{MoonEm}

Since MoonEm is a constant-delay emulator, we evaluate it only for a jitter of \qty{0}{\ms}.
We also configure MoonEm without bandwidth limitation or packet loss.
MoonEm's authors report NIC-dependent forwarding rates (Intel~E810 vs. NVIDIA~ConnectX-5).
Our measurement therefore serves as a baseline for DPDK-based emulators on the ConnectX-5 NIC.
We measured a maximum forwarding rate of MoonEm of \qty{46}{\gbps}, approximately 9 times NetEm's rate.
This shows that the kernel bypass approach outperforms the in-kernel approach of NetEm.

\paragraph{DPDS}

The maximum forwarding rate of \ac{DPDS} using reordering is \qty{78.5}{\gbps} and \qty{75.8}{\gbps} for \qty{0}{\ms} and \qty{3}{\ms} jitter, respectively.
By contrast, adaptive delay correlation reaches \qty{95.5}{\gbps} and \qty{93.6}{\gbps} for the same jitters, which is approximately $21.6\%$ and $23.5\%$ higher.
The difference is due to the sorting overhead of reordering, which is avoided by adaptive delay correlation.
For a jitter of \qty{3}{\ms}, spacing slightly raises the maximum forwarding rate for adaptive delay correlation (from \qty{93.6}{\gbps} to \qty{94.9}{\gbps}).
This is because spacing prevents the \ac{NIC} from being overloaded (see \autoref{subsubsec:increasingDPDSZLT}).
Thus, correlated varying delays are more efficient for high-rate emulation than reordering.

\ac{DPDS} with packet reordering implements the same strategy as NetEm.
However, \ac{DPDS}'s maximum forwarding rate using reordering is 15.4 times that of NetEm and approximately 1.7 times that of MoonEm.
As with MoonEm, we attribute this to the different implementation approaches: in-kernel versus kernel bypass.
Furthermore, \ac{DPDS}'s maximum forwarding rate with adaptive delay correlation is 18.7 times that of NetEm and over twice that of MoonEm.
Overall, \ac{DPDS} outperforms both baselines through a combination of its kernel bypass implementation and adaptive delay correlation, a reordering-free emulation strategy.

\subsubsection{Zero-Loss Throughput}
\label{subsubsec:increasingDPDSZLT}

First, we compare the measured \ac{ZLT} of NetEm, MoonEm, and \ac{DPDS} in different configurations.
Second, we explain why spacing increases the \ac{ZLT} of adaptive delay correlation.
For NetEm and MoonEm, we observe a transient phase with packet loss at the beginning of each measurement.
The reported \ac{ZLT} therefore refers to the steady-state throughput after this phase without further packet loss.
\ac{DPDS} does not exhibit such a transient phase.

\renewcommand{\cellalign}{cc}
\begin{table}[t]
  \caption{Measured \acl{ZLT} of NetEm, MoonEm, and DPDS.}
  \label{tab:zlt}
  \centering
  \begin{tabularx}{\columnwidth}{l|YY}
    \toprule
                          & \multicolumn{2}{c}{\emph{Jitter}}                       \\
                          & \emph{\qty{0}{\ms}}               & \emph{\qty{3}{\ms}} \\
    \midrule
    NetEm                 & \qty{5.0}{\gbps}                  & \qty{4.5}{\gbps}    \\
    NetEm + TBF           & \qty{4.6}{\gbps}                  & \qty{4.2}{\gbps}    \\
    MoonEm                & \qty{44.8}{\gbps}                 & $\times$            \\
    DPDS (reordered)      & \qty{73.5}{\gbps}                 & \qty{58.8}{\gbps}   \\
    DPDS (corr.)          & \qty{95.0}{\gbps}                 & \qty{60.8}{\gbps}   \\
    DPDS (corr. + spaced) & \qty{94.4}{\gbps}                 & \qty{85.2}{\gbps}   \\
    \bottomrule
  \end{tabularx}
\end{table}

\paragraph{Comparison of DPDS with NetEm and MoonEm}

\autoref{tab:zlt} shows that \ac{DPDS} achieves substantially higher \acp{ZLT} than NetEm and MoonEm across all jitter configurations.
For a jitter of \qty{0}{\ms}, \ac{DPDS} with adaptive delay correlation reaches a \ac{ZLT} of \qty{95.0}{\gbps}, 19 times that of NetEm and over 2 times that of MoonEm.
At a jitter of \qty{3}{\ms}, \ac{DPDS} with adaptive delay correlation reaches \qty{60.8}{\gbps}, still over 13 times that of NetEm.
MoonEm does not support varying delays and is therefore omitted at this jitter.
\ac{DPDS} with reordering achieves \qty{73.5}{\gbps} and \qty{58.8}{\gbps} for jitters of \qty{0}{\ms} and \qty{3}{\ms}, respectively, also clearly exceeding NetEm and MoonEm.
Adding \ac{TBF} to NetEm decreases its \ac{ZLT} slightly due to the additional processing overhead.
However, at NetEm's low rates, the burst-smoothing benefit of spacing observed for \ac{DPDS} does not materialize.

\paragraph{Effect of Spacing on Adaptive Delay Correlation's Zero-Loss Throughput}

For constant delays (jitter of \qty{0}{\ms}), spacing slightly decreases \ac{DPDS}'s (corr.) \ac{ZLT} from \qty{95.0}{\gbps} to \qty{94.4}{\gbps} due to the additional processing overhead.
For varying delays (jitter of \qty{3}{\ms}), however, spacing increases the \ac{ZLT} from \qty{60.8}{\gbps} to \qty{85.2}{\gbps}, an improvement of approximately $40\%$.
This practically confirms an effect already observed in the spacing simulation (see \autoref{subsubsec:adaptiveCorrelationWithSpacing}): adaptive delay correlation inherently produces packet bursts when successive delays are similar.
At low rates these bursts fit within the \ac{NIC}'s buffer, but at high rates they exceed its size and cause packet loss.
Applying spacing before the \ac{NIC} receives the packets from the emulator smooths these bursts so they fit within the available buffer, preventing such overload.
Therefore, despite the small overhead in the constant-delay case, spacing substantially increases \ac{DPDS}'s (corr.) \ac{ZLT} for varying delays.


\subsection{Accuracy of Delay Strategies}
\label{subsec:accuracyVaryingDelay}

\begin{figure}[t]
  \centering
  \subfloat[Mean delay accuracy.\label{fig:emulator_delay}]{%
    \includegraphics[width=\columnwidth]{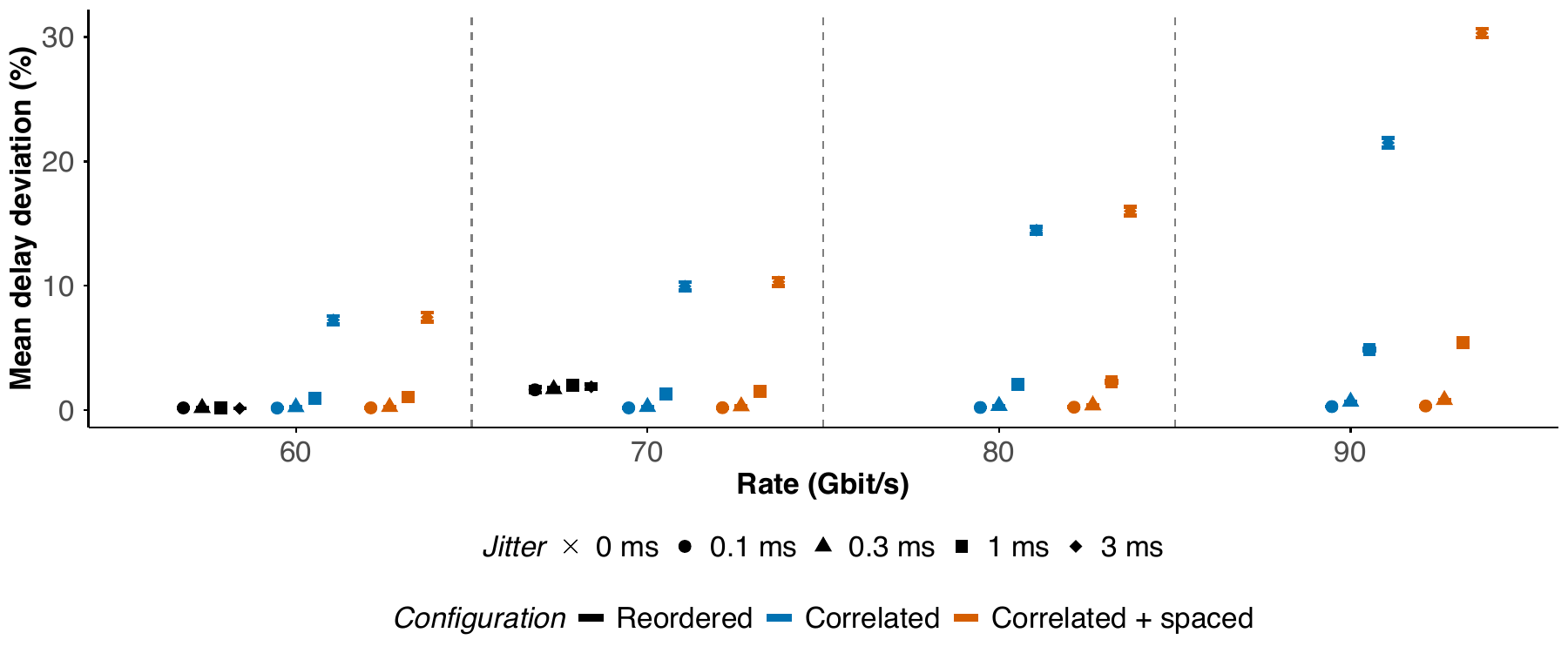}%
  }\\
  \subfloat[Jitter accuracy.\label{fig:emulator_jitter}]{%
    \includegraphics[width=\columnwidth]{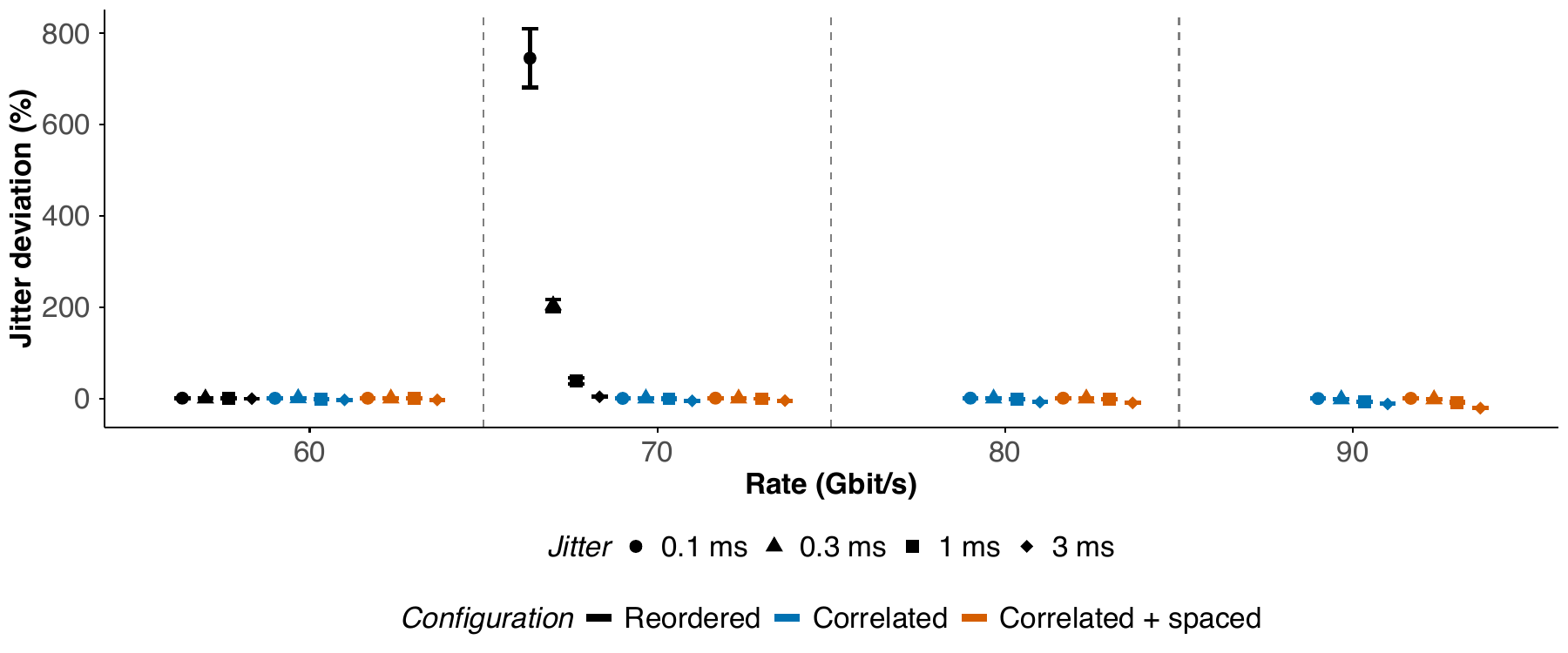}%
  }
  \caption{Measured accuracy of \ac{DPDS} for reordered and correlated ($t_h = \qty{21}{\ms}$, with and without spacing) delays at a desired mean delay of \qty{10}{\ms} across various rates and jitters.}
  \label{fig:emulatorEval}
\end{figure}

We evaluate the accuracy of the \ac{DPDS} prototype for reordering and adaptive delay correlation at rates not exceeding their respective maximum forwarding rates.
\autoref{fig:emulatorEval} shows the accuracy of the \ac{DPDS} prototype for different combinations of emulation configuration, rate, and jitter.
The measured accuracies confirm the simulation results from \autoref{subsec:comparingAccuracy}.

\subsubsection{Accuracy with Reordering}

\autoref{fig:emulator_delay} shows the mean delay deviations for rates below the maximum forwarding rate with \ac{DPDS} and reordering.
The deviations of reordering are greater than those observed in the simulation.
This is because \ac{DPDS} processes up to 16 packets at a time, resulting in packet bursts.
Additionally, the \ac{NIC} enforces a minimum \ac{IAT} between successive packets.
Together, these effects prevent \ac{DPDS} with reordering from realizing the fine-grained delays observed in simulation.

For jitters of \qty{0.1}{\ms} and \qty{0.3}{\ms} at \qty{70}{\gbps}, the jitter deviations exceed those of the simulations (see \autoref{fig:emulator_jitter}).
This is due to \ac{DPDS}'s batch processing and the minimum \ac{IAT} enforced by the \ac{NIC}, both of which become more pronounced as the traffic rate approaches the \ac{ZLT} of \ac{DPDS} with reordering.
At small desired jitters, even modest absolute deviations translate into large relative deviations, amplifying the visible error.

\subsubsection{Accuracy with Adaptive Delay Correlation}

\autoref{fig:emulatorEval} shows that for correlation, the mean delay and jitter deviations match those observed in the simulation for limited bandwidth (see \autoref{subsubsec:adaptiveCorrelationWithSpacing}).
As with reordering, the deviations are slightly increased due to \ac{DPDS}'s batch processing.
In some configurations, such as a jitter of \qty{0.1}{\ms}, adaptive delay correlation is even more accurate than reordering, suggesting that it is less sensitive to batch processing.
The deviations further increase as the traffic rate approaches the adaptive delay correlation's \ac{ZLT}.

\autoref{fig:emulatorEval} shows that \ac{DPDS}'s additional spacing has a negligible effect on accuracy, since the \ac{NIC} already spaces the packets to its bandwidth.
The only exception occurs for a jitter of \qty{3}{\ms} at \qty{90}{\gbps}.
Although \ac{DPDS} spaces the packets, the packets still arrive at the \ac{NIC} in bursts due to the maximum burst size of 16 packets.
The \ac{NIC} therefore has to smooth these bursts itself, which adds queuing delay and increases the realized mean delay.


\subsection{Adaptation to a Changing Rate}
\label{subsec:measuredAccuracyChangingRate}

We show that adaptive delay correlation maintains accurate delay emulation even under changing traffic rates.

\subsubsection{Configuration of a Changing Rate}

To evaluate \ac{DPDS} under a changing rate, we configure P4TG to generate a sinusoidal traffic pattern.
The sinusoidal shape continuously exposes the emulator to both increasing and decreasing rates, which stresses the rate estimation in both directions.
The pattern has a maximum rate of \qty{50}{\gbps} and a period of \qty{30}{\s}.
Over a measurement duration of \qty{4}{\min}, the traffic therefore goes through eight full periods, with the rate changing by approximately \qty{3.3}{Gbit/s^2} on average between its minimum and maximum.
This rate of change is fast enough to challenge the rate estimation but slow enough that each rate remains representative for a meaningful number of packets.

\subsubsection{Accuracy at a Changing Rate}

\autoref{tab:sineDeviations} lists the mean delay and jitter deviations at the end of each experiment.
Further, it includes the deviations for \ac{CBR} traffic at \qty{50}{\gbps} as a reference.
Overall, the deviations for both traffic patterns are within the same range and grow with increasing jitter.
For the mean delay, the sinusoidal pattern consistently deviates less than \ac{CBR}, while neither pattern is consistently better than the other for the jitter.
Note that the sinusoidal pattern has a lower average packet rate than \ac{CBR} at \qty{50}{\gbps}, putting less load on the emulator.
The similar deviations therefore primarily indicate that adaptive delay correlation's rate estimation works as intended.
This confirms that adaptive delay correlation with a rate-dependent \ac{EMA} is suitable for emulating varying delays at changing traffic rates.

\renewcommand{\cellalign}{cc}
\begin{table}[t]
  \caption{Measured accuracy of \ac{DPDS}'s adaptive delay correlation ($t_h = \qty{21}{\ms}$) at a \ac{CBR} and sinusoidal traffic pattern at a desired mean delay of \qty{10}{\ms} across various jitters.}
  \label{tab:sineDeviations}
  \centering
  \begin{tabularx}{\columnwidth}{r|YY}
    \toprule
    \emph{Jitter}  & \emph{CBR}                                                                   & \emph{Sine}                                                                  \\
    \midrule
                   & \multicolumn{2}{c}{\emph{Mean delay deviation}}                                                                                                             \\
    \qty{0.1}{\ms} & \qty[separate-uncertainty, retain-zero-uncertainty]{0.33 +- 0.00}{\percent}  & \qty[separate-uncertainty, retain-zero-uncertainty]{0.29 +- 0.00}{\percent}  \\
    \qty{0.3}{\ms} & \qty[separate-uncertainty, retain-zero-uncertainty]{0.52 +- 0.02}{\percent}  & \qty[separate-uncertainty, retain-zero-uncertainty]{0.47 +- 0.01}{\percent}  \\
    \qty{1}{\ms}   & \qty[separate-uncertainty, retain-zero-uncertainty]{1.17 +- 0.07}{\percent}  & \qty[separate-uncertainty, retain-zero-uncertainty]{1.10 +- 0.13}{\percent}  \\
    \qty{3}{\ms}   & \qty[separate-uncertainty, retain-zero-uncertainty]{2.97 +- 0.44}{\percent}  & \qty[separate-uncertainty, retain-zero-uncertainty]{2.74 +- 0.49}{\percent}  \\
    \midrule
                   & \multicolumn{2}{c}{\emph{Jitter deviation}}                                                                                                                 \\
    \qty{0.1}{\ms} & \qty[separate-uncertainty, retain-zero-uncertainty]{0.23 +- 0.13}{\percent}  & \qty[separate-uncertainty, retain-zero-uncertainty]{0.64 +- 0.20}{\percent}  \\
    \qty{0.3}{\ms} & \qty[separate-uncertainty, retain-zero-uncertainty]{-0.45 +- 0.25}{\percent} & \qty[separate-uncertainty, retain-zero-uncertainty]{-0.19 +- 0.22}{\percent} \\
    \qty{1}{\ms}   & \qty[separate-uncertainty, retain-zero-uncertainty]{-0.49 +- 0.29}{\percent} & \qty[separate-uncertainty, retain-zero-uncertainty]{-0.20 +- 0.37}{\percent} \\
    \qty{3}{\ms}   & \qty[separate-uncertainty, retain-zero-uncertainty]{-0.55 +- 0.55}{\percent} & \qty[separate-uncertainty, retain-zero-uncertainty]{-0.88 +- 0.98}{\percent} \\
    \bottomrule
  \end{tabularx}
\end{table}

%% file: content/conclusions.tex
\section{Conclusions}
\label{sec:conclusions}

In this paper, we tackled the problem of adding varying delay to packets for link emulation.
The challenge is that varying delay may add more delay to packets than intended when preceding packets are extensively delayed.
Packet reordering solves this problem, but is detrimental to many applications and should be avoided.
We proposed adaptive delay correlation to largely mitigate this problem.
It essentially generates independent delay values and smooths them with an \ac{EMA}.
The algorithm takes a desired mean delay and standard deviation (jitter) as input, as well as a half-life period to control delay dynamics over time.
Short-term rate measurements adapt the \ac{EMA}'s weight to different traffic rates.

We investigated adaptive delay correlation by simulation and showed that a half-life period of \qty{21}{\ms} achieves mean delay and jitter as desired while avoiding overly slow delay dynamics.
Bandwidth limitation deteriorates the accuracy of mean delay and jitter at high utilization.
These problems are avoided with packet reordering, but the reordering effect is substantial and undesired.

We further developed a \acf{DPDS}, which is a kernel bypass approach for link emulation.
It implements adaptive delay correlation, packet reordering, spacing, policing, and two models for packet loss.
Our performance evaluation showed that \ac{DPDS} achieves a \ac{ZLT} of \qty{95}{\gbps} for constant delay and, with spacing enabled, \qty{85}{\gbps} for varying delay with a jitter of \qty{3}{\ms} while meeting the desired delay.
With these values, \ac{DPDS} clearly outperforms the widely used \ac{NLE} NetEm and the recently developed DPDK-based emulator MoonEm.
With packet reordering, \ac{DPDS} achieves \acp{ZLT} of only \qty{73}{\gbps} for constant and \qty{58}{\gbps} for varying delay, which still exceed those of NetEm and MoonEm.
As adaptive delay correlation relies on continuous rate measurement, we also showed that \ac{DPDS} works as desired under changing rates.
\ac{DPDS} is open source and available on GitHub~\cite{GitHubDPDS}.

%% file: content/authors.tex
\begin{IEEEbiography}[{\includegraphics[width=1in,height=1.25in,clip,keepaspectratio]{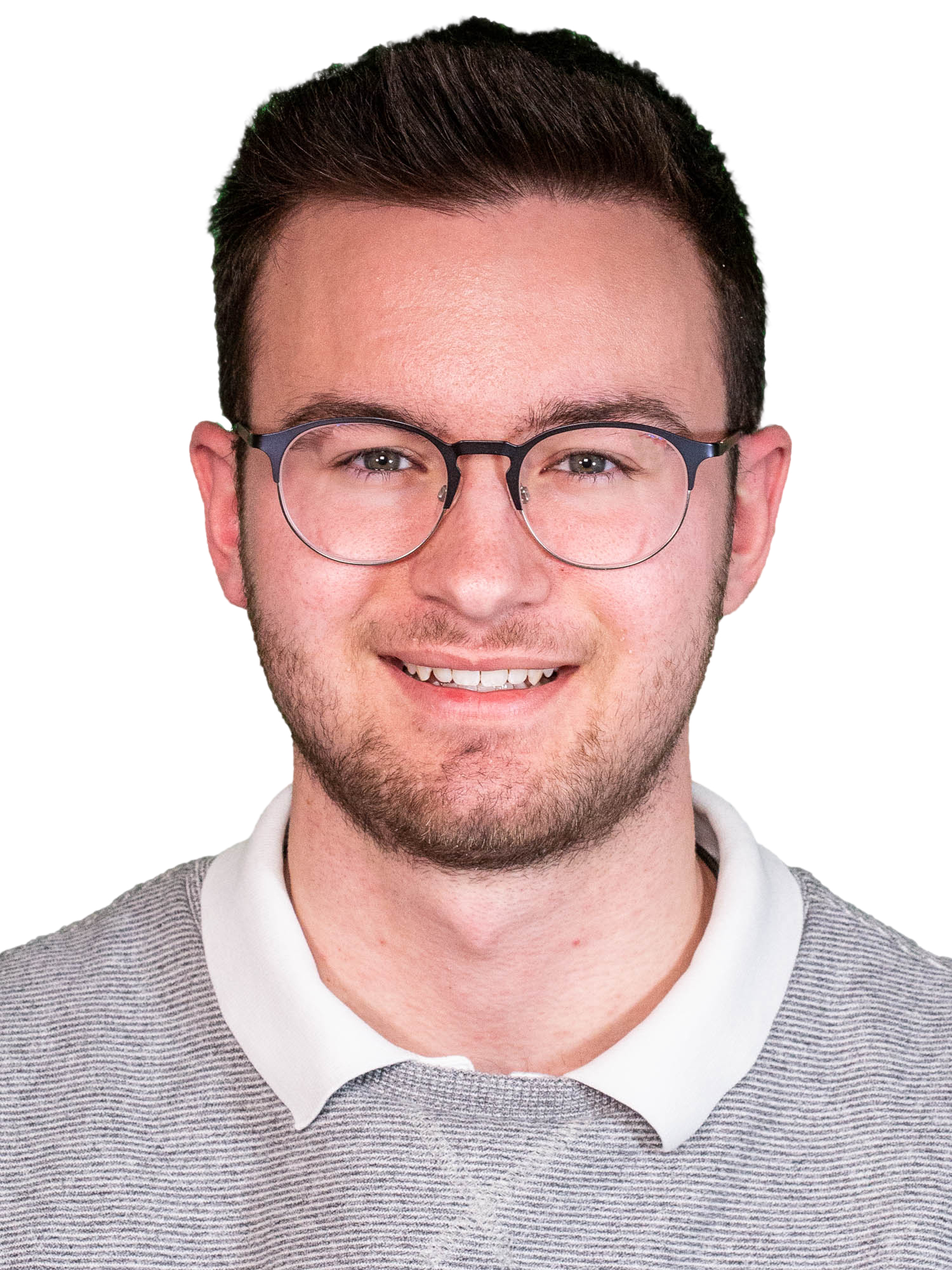}}]{Etienne Zink}
received his bachelor's degree (2022) from the Corporate State University Baden-W\"urttemberg and his master's degree (2024) from the University of T\"ubingen, both in computer science.
Afterward, he joined the communication networks research group of Prof. Dr. habil. Michael Menth as a Ph.D. student.
His research interests include software-defined networking, network function virtualization, resilience, and network emulation.
\end{IEEEbiography}

\begin{IEEEbiography}[{\includegraphics[width=1in,height=1.25in,clip,keepaspectratio]{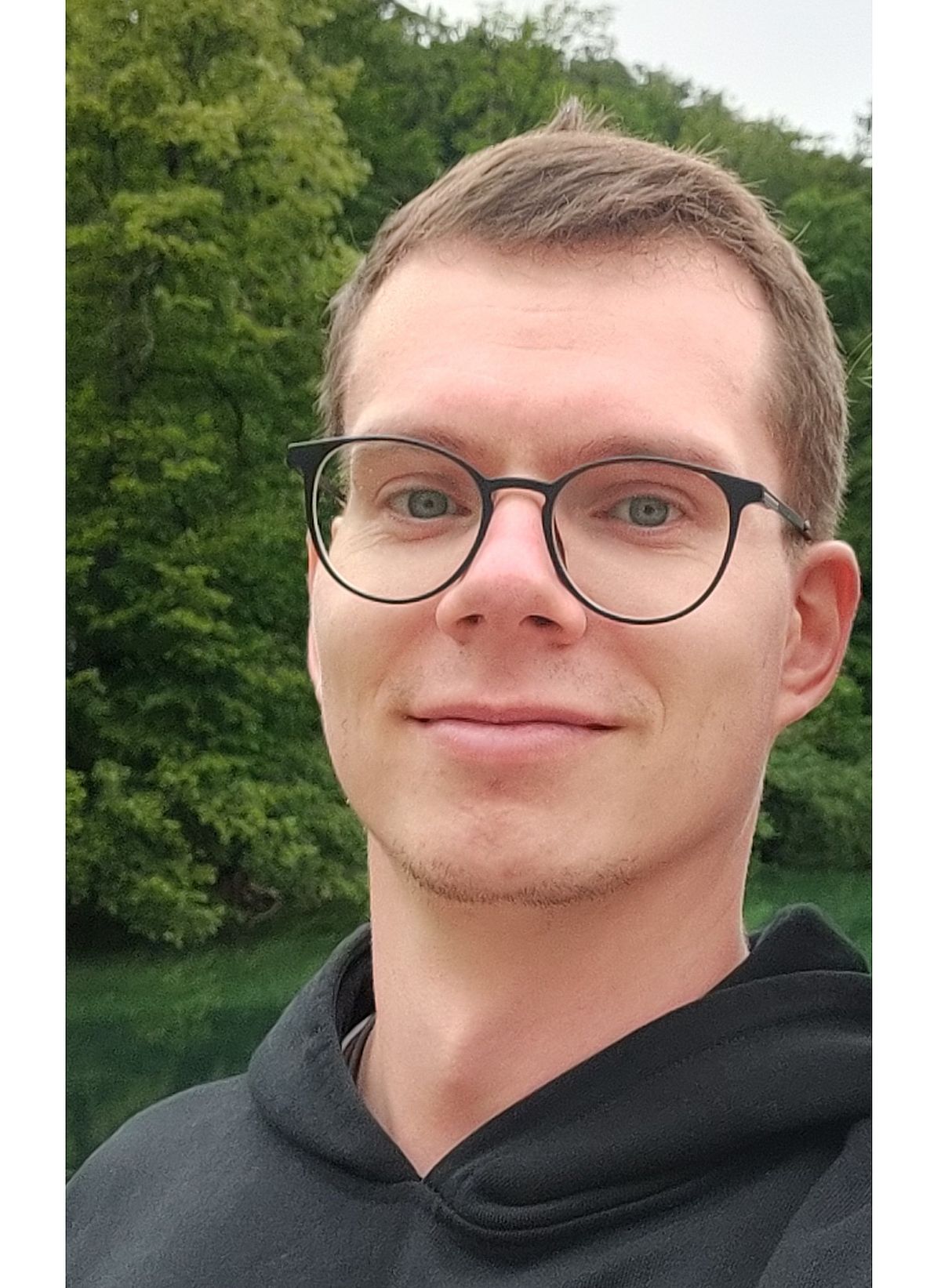}}]{Fabian Ihle}
received his bachelor's (2021) and master's degrees (2023) in computer science from the University of T\"ubingen.
Afterward, he joined the communication networks research group of Prof. Dr. habil. Michael Menth as a Ph.D. student.
His research interests include software-defined networking, P4-based data plane programming, resilience, and Time-Sensitive Networking (TSN).
\end{IEEEbiography}

\begin{IEEEbiography}[{\includegraphics[width=1in,height=1.25in,clip,keepaspectratio]{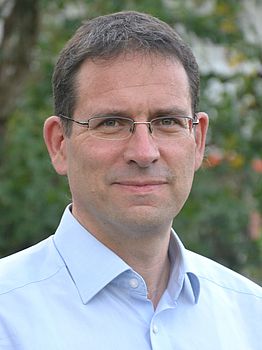}}]{Michael Menth}
is a professor at the Department of Computer Science at the University of T\"ubingen/Germany and chairholder of Communication Networks since 2010.
He studied, worked, and obtained diploma (1998), PhD (2004), and habilitation (2010) degrees at the universities of Austin/Texas, Ulm/Germany, and W\"urzburg/Germany.
His special interests are performance analysis and optimization of communication networks, resilience and routing issues, as well as resource and congestion management.
His recent research focus is on network softwarization, in particular P4-based data plane programming, Time-Sensitive Networking (TSN), Internet of Things, and Internet protocols.
Dr. Menth contributes to standardization bodies, notably to the IETF.
\end{IEEEbiography}